\documentclass{aa}

\usepackage{graphicx}
\usepackage{txfonts}
\usepackage{natbib}

\newcommand\bb[1] {   \mbox{\boldmath{$#1$}}  }
\newcommand\del{\bb{\nabla}}
\newcommand\bcdot{\bb{\cdot}}
\newcommand\btimes{\bb{\times}}

\begin{document}

\title{Turbulent resistivity driven by the magnetorotational instability}
\author{S\'ebastien Fromang \inst{1,2} and James M. Stone \inst{3}}

\offprints{S.Fromang}

\institute{CEA, Irfu, SAp, Centre de Saclay, F-91191 Gif-sur-Yvette, F
rance \and UMR AIM, CEA-CNRS-Univ. Paris VII, Centre de Saclay, F-91191
Gif-sur-Yvette, France. \and Department of Astrophysical Sciences,
Peyton Hall, Ivy Lane, Princeton University, NJ 08544, USA. \\ \email{sebastien.fromang@cea.fr}}

\date{Accepted; Received; in original form;}

\label{firstpage}

\abstract
{}
{We measure the turbulent resistivity in the nonlinear regime of the
MRI, and evaluate the turbulent magnetic Prandtl number.}
{We perform a set of numerical simulations with the Eulerian finite
  volume codes Athena and Ramses in the framework of the shearing
  box model. We consider models including explicit dissipation
  coefficients and magnetic field topologies such that the net
  magnetic flux threading the box in both the vertical and 
  azimuthal directions vanishes.}
{We first demonstrate good agreement between the two codes by
comparing the properties of the turbulent states in
simulations having 
identical microscopic diffusion coefficients (viscosity and
resistivity).  We find the properties of the turbulence do
not change when the box size is increased in the radial direction,
provided it is elongated in the azimuthal direction. To measure
the turbulent resistivity in the disk, we impose a fixed
electromotive force on the flow
and measure the amplitude of the saturated magnetic field
that results. We obtain a turbulent resistivity that is in rough
agreement with mean field theories like the Second Order Smoothing
Approximation.  The numerical value translates into a turbulent magnetic
Prandtl number 
$Pm_t$ of order unity. $Pm_t$ appears to be an increasing function of
the forcing we impose. It also becomes smaller as the box size is
increased in the radial direction, in good agreement with previous
results obtained in very large boxes.}
{Our results are in general agreement with other recently published papers
  studying the same problem but using different methodology.  Thus, our
conclusion that $Pm_t$ is of order unity
  appears robust.}
\keywords{Accretion, accretion discs - MHD - Methods: numerical}

\authorrunning{S.Fromang \&  J.Stone}
\titlerunning{Turbulent resistivity driven by the MRI}
\maketitle

\section{Introduction}

Magnetic fields play a central role in accretion
disk theory.  Through the magnetorotational instability
\citep[MRI,][]{balbus&hawley91,balbus&hawley98}, they can explain why
the flow is turbulent and how angular momentum
is transported outward.
They are also thought to be responsible for jet launching and
collimation \citep{blandford&payne82,pudritzetal07}. For both phenomena,
whether or not a mean magnetic flux threads the disk has important
consequences. Numerical simulations have indeed shown that the saturation
level of the MRI depends strongly on the net flux: the famous $\alpha$
parameter \citep[a measure of
  the rate of angular momentum transport through the disk,
see][]{shakura&sunyaev73} is directly proportional to the square of
the vertical magnetic field \citep{hawleyetal95}. Similarly, a strong
vertical field is mandatory for accretion--ejection models to be efficient
at launching jets \citep{casse&ferreira00}.

However, MHD turbulence not only can transport angular
momentum outward in the disk, but also can diffuse
magnetic flux radially. 
This possibility questions the very
presence of a mean magnetic flux in the inner part of accretion disks
\citep{vanballegooijen89,lubowetal94}. Ultimately, how much magnetic
flux resides in the inner disk depends on a competition between the
effectiveness of outward angular momentum transport (i.e. how well
mass and magnetic flux are advected inward) and magnetic flux
diffusion. The former effect can be identified with a ``turbulent
viscosity''. Similarly, the importance of the latter can be assessed
through an equivalent ``turbulent resistivity'' $\eta_t$
which has physical consequences similar to that of a microscopic
resistivity. The purpose of this paper is to measure $\eta_t$ accurately
by using a set of local numerical simulations in which the flow is
turbulent because of the MRI.

We note that two recently published papers have considered the same
problem \citep{guan&gammie09,lesur&longaretti09}. Although all such
studies (including our own) have used numerical simulations in the
local shearing box model, there are significant differences between
the approach used in each case.  First, the numerical methods used
to solve the MHD equations are different in all three approaches:
\citet{guan&gammie09} used numerical methods similar to those
in the ZEUS code \citep{stone&norman92b,hawley&stone95} and
\citet{lesur&longaretti09} used a pseudo-spectral code in the
incompressible limit. By contrast, we use two finite volume codes,
Athena \citep{gardiner&stone05a, gardiner&stone2008,stoneetal08}
and Ramses \citep{Teyssier02,Fromangetal06}. In addition, the
method used in this paper to measure the turbulent resistivity (see
section~\ref{setup}) is different from that used by \citet{guan&gammie09}
and \citet{lesur&longaretti09}. The former add a large scale magnetic
field to an already turbulent flow and associate the rate of decay of this
field to the turbulent resistivity in the disk. The latter impose at all
times an additional magnetic field in the computational domain, measure
the time averaged electromotive force (EMF) associated with this field,
and use the amplitude of the EMF to measure the turbulent resistivity.
As described below, our approach is to impose a forcing EMF on the flow
and relate the time averaged structure of the saturated field that results
to the value $\eta_t$. Another difference between our work and previous
studies is that we study turbulence driven by a magnetic field with not
net flux threading the box in any direction, while \citet{guan&gammie09}
and \citet{lesur&longaretti09} perform numerical simulations in the
presence of either a net vertical or a net toroidal field. It is well
known that the presence of a net field affects the properties of the
turbulence that develops \citep{hawleyetal95,guanetal09}. Thus, an
additional goal of this paper will be to study how much the final results
obtained in all these studies depends on the details of the simulation
setup. This will be useful to assess the robustness of the results.

The plan of this paper is as follows. In section~\ref{setup}, we detail
our method to measure the turbulent resistivity and the numerical
setup we used. Section~\ref{noforcing_sec} serves essentially to
validate our numerical codes by considering the case of a vanishing
forcing EMF (in which case the simulations are standard shearing box
simulations). We show that we obtain turbulent flows whose time- and
volume-averaged properties are identical with both Athena and Ramses.
We also study whether these results are modified when we consider
boxes extended in the radial direction, as suggested recently in the
literature \citep{johansenetal09}. These simulations serve as a basis for
section~\ref{forcing_sec} in which we measure the turbulent resistivity in
a number of situations, varying the strength and direction of the forcing
EMFs, and the size of the box. Finally, in section~\ref{conclusion_sec},
we summarize our results, compare them with those of \citet{guan&gammie09}
and \citet{lesur&longaretti09}, and stress the limitations of our study.

\section{The setup}
\label{setup}

\subsection{Equations}

We solve the equations of magnetohydrodynamics (MHD) in the
framework of the shearing box model \citep{goldreich&lyndenbell65}
using two different codes: Athena \citep{gardiner&stone05a,
gardiner&stone2008,stoneetal08}
and Ramses \citep{Teyssier02,Fromangetal06}.  Explicit microscopic
dissipation, namely viscosity $\nu$ and resistivity $\eta$, are included
in the simulations (note the microscopic resistivity $\eta$
should not be confused with the turbulent resistivity $\eta_t$ that will be
introduced below). It has been shown in the last few years that both are
important in setting the saturated state of turbulence driven by the MRI
\citep{lesur&longaretti07,fromangetal07}.  We use an isothermal equation
of state $P=\rho c_0^2$ to relate density $\rho$ and pressure $P$, where
$c_0$ is the sound speed. In a Cartesian coordinate system $(x,y,z)$
with unit vectors $(\bb{i},\bb{j},\bb{k})$ in
the radial, azimuthal and vertical directions respectively,
the equations for mass and momentum conservation are:
\begin{eqnarray}
\frac{\partial \rho}{\partial t} + \del \bcdot (\rho \bb{v})  &=&  0
\,\label{contg} \label{mass_eq} \, , \\ 
\frac{\partial \rho \bb{v}}{\partial t} + \del \bcdot (\rho \bb{v}
\bb{v} - \bb{B} \bb{B}) + \del P_{tot} &=& 
 3\rho \Omega x \bb{i} - 2 \rho \bb{\Omega} \times \bb{v} + \nabla
 \bcdot \bb{T} \label{momentum_eq} \, ,
\end{eqnarray}
where $\Omega$ is the angular velocity of the shearing box around the
central object, $\bb{B}$ is the magnetic field, $P_{tot}=P+(\bb{B}
\bcdot \bb{B})/2$ is the total pressure, and $\bb{T}$ is the viscous
stress tensor \citep{landau&lifchitz59}. The first two source terms
on the right hand side of Eq.~(\ref{momentum_eq}) represent
the tidal gravity and Coriolis forces.  Their
implementation in Athena and Ramses is not straightforward.  We
followed the method described by \citet{gardiner&stone05b} 
(see also \citet{stone&gardiner09}) and use a
Crank--Nicholson algorithm to update the momentum fluctuations in time
due to these source terms.
The divergence of the viscous stress tensor in Eq.~(\ref{momentum_eq})
is differenced in the conservative
form so as to exactly conserve total momentum.

In order to study turbulent resistivity, we use a
modified form of the induction equation:
\begin{equation}
\frac{\partial \bb{B}}{\partial t}  =  \del \btimes \left( \bb{v} \btimes
\bb{B}  - \eta \del \btimes \bb{B} + \bb{E} \right) \, \label{induct} .
\label{induction_eq}
\end{equation} 
In this form we have introduced $\bb{E}$, an imposed electromotive force
(EMF), in addition to the usual inductive and resistive terms.  When
$\bb{E}=0$, as in section~\ref{noforcing_sec}, we recover the usual
form of the induction equation for resistive MHD.
In section~\ref{forcing_sec}, we consider solutions to
Eqs.~(\ref{mass_eq})--(\ref{induction_eq}) using the following expression for
the imposed EMF: 
\begin{equation}
\bb{E}=E_{0y} \cos \left(\frac{2\pi x}{\lambda_{E,x}} \right) \bb{j} + E_{0x}
\cos \left(\frac{2\pi z}{\lambda_{E,z}} \right) \bb{i} \, ,
\end{equation} 
where $\lambda_{E,x}$ and $\lambda_{E,z}$ are the radial
and vertical wavelengths for the forcing EMF, respectively.
To gain insight into the effect of $\bb{E}$, it is instructive to
consider the case $\bb{v}=0$. In this situation, the magnetic field
reaches a steady state in which forcing due to $\bb{E}$ is balanced by
resistive dissipation, with the field given by the following solution
\begin{equation}
\bb{B}=-\frac{E_{0x}\lambda_{E,z}}{2 \pi \eta} \sin \left( \frac{2\pi z}{\lambda_{E,z}}
\right) \bb{j} - \frac{E_{0y} \lambda_{E,x}}{2 \pi \eta} \sin \left(\frac{2\pi
  x}{\lambda_{E,x}}\right) \bb{k} \, . 
\label{steady_b}
\end{equation}
Eq.~(\ref{steady_b}) shows that the effect of the radial (x) component of
the EMF is to create a vertically varying azimuthal magnetic field,
while the effect of the azimuthal (y) component of the
EMF is to create a radially varying vertical magnetic
field. The amplitude of the field is determined by the amplitudes and
wavelengths of the forcing and the resistivity.  In a turbulent flow
we expect the same balance will be achieved: turbulent velocity
fluctuations will diffuse and reconnect the magnetic field that the driving
EMF is trying to build. By analogy with Eq.~(\ref{steady_b}), in a turbulent 
flow, we can define a turbulent resistivity  $\eta_{t}$ by measuring the
amplitude of the spatially varying field resulting from the
forcing.  More specifically, we will consider two cases. First,
$E_{0x}=0$ and $E_{0x}>0$, in which case the forcing results in a
purely vertical field of amplitude $B_z^0$. This allows us
to measure the turbulent resistivity $\eta_{t,x}$ in the radial direction
through the following relation, obtained using Eq.~(\ref{steady_b}), 
\begin{equation}
\eta_{t,x}=\frac{E_{0y}}{2\pi}\frac{\lambda_{E,x}}{B_z^0} \, .
\label{eta_turb_x}
\end{equation}
Similarly, the case $E_{0x}>0$ and $E_{0x}=0$ yields a purely
azimuthal field varying with z and with an amplitude $B_y^0$ that 
measures the turbulent resistivity $\eta_{t,z}$ in the
vertical direction according to
\begin{equation}
\eta_{t,z}=\frac{E_{0x}}{2\pi}\frac{\lambda_{E,z}}{B_y^0} \, .
\label{eta_turb_z}
\end{equation}

\subsection{Numerical method}

We solved Eq.~(\ref{mass_eq}), (\ref{momentum_eq}) and
(\ref{induction_eq}) using Athena and Ramses in a box of size
$(L_x,L_y,L_z)=(H,4H,H)$, where $H=c_0/\Omega$ defines the vertical
scale height (thickness) of the disk.
We also used Athena to compute a few runs with a larger
extent in the radial direction:
$(L_x,L_y,L_z)=(4H,4H,H)$.  It has 
been shown recently by \citet{johnsonetal08} that truncation error
in the shearing box
can create numerical artifacts, such as a density minima at the
center of the box (where the shear velocity with respect to the grid is zero).
We have implemented an orbital advection algorithm
\citep{masset00,gammie01,johnson&gammie05,johnsonetal08} in Athena
\citep{stone&gardiner09} to alleviate
this problem, and at the same time to allow for much larger timesteps
in wide radial boxes.

All of our simulations are performed with no net magnetic flux
in either the vertical or azimuthal direction. At $t=0$, we
start with a purely vertical magnetic field varying sinusoidally with
$x$ so that the mean field vanishes. Random velocity perturbations
of small amplitude are applied to the background state, consisting of
a uniform density gas with a linear shear profile: $\bb{v}=(0,-3\Omega
x/2,0)$.  We adopt $\Omega=c_0=10^{-3}$. In the rest of this
paper, we measure time in units of the 
orbital period $T_{orb}=2\pi/\Omega$. In all of the simulations we
present below, we set $\nu$ and $\eta$ such that the 
Reynolds number $Re=c_0H/\eta=3125$ and the magnetic Prandtl number
$Pm=4$. These coefficients are identical to the run labeled
{\it 128Re3125Pm4} of \citet{fromangetal07}, and are known to lead to
sustained turbulence over large numbers of orbital times. For explicit
dissipation to dominate over numerical dissipation, we
used a resolution of 128 grid points per $H$, or $(Nx,Ny,Nz)=(128,512,128)$
for calculations that span one scale height in the radial direction.
Such a resolution has been shown to give resolved solutions 
when using 2nd order Eulerian codes like ZEUS 
\citep{fromangetal07} for microscopic dissipation coefficients
of the magnitude adopted here. \citet{simonetal09} have recently
shown that the numerical dissipation in Athena, likely comparable
to Ramses, is similar in magnitude to that of ZEUS 
when performing numerical simulations of MRI--induced MHD turbulence
in the shearing box. Thus, we expect the same resolution needed
with ZEUS is also appropriate for the runs presented here that use
Athena or Ramses. In simulations with 
larger boxes, we scaled the resolution accordingly,
$(Nx,Ny,Nz)=(512,512,128)$, so that the cell size remains the same.

Regardless of the code we use or the size of the computational domain,
to study the turbulent resistivity induced by the MRI
our strategy is as follows: we first performed a run without forcing
(i.e. $\bb{E}=0$) to 
let the MRI develop and for turbulence to reach a steady state.  Typically
these runs
are evolved for $100$ orbits.  They also serve as a useful
comparison between codes, and to compare small vs. large boxes. At $t=30$,
we restarted the simulations with forcing of various amplitudes imposed,
as described above.  With forcing, the flow evolves toward a new quasi
steady--state that we use to evaluate the turbulent resistivity
induced by the MRI. Before moving on to a description of the results with
forcing in section~\ref{forcing_sec}, we first describe the properties of MHD
turbulence as obtained in runs without forcing in the next
section. 

\section{Runs without forcing}
\label{noforcing_sec}

\begin{table*}[t]
\begin{center}
\begin{tabular}{@{}cccccccc}\hline\hline
Model & Box size & Resolution & t$_{\textrm{lim}}$ &
$\alpha$ & $\delta v_x/c_0$ &
$\delta v_y/c_0$ & $\delta v_z/c_0$  \\
\hline\hline
R--SB & $(H,4H,H)$ & $(128,512,128)$ & $160$ & $1.4 \times
10^{-2} \pm 3.6 \times 10^{-3}$ & 0.0982 & 0.0776 & 0.0609 \\
A--SB & $(H,4H,H)$ & $(128,512,128)$ & $100$ & $1.5 \times
10^{-2} \pm 5.7 \times 10^{-3}$ & 0.0936 & 0.0774 & 0.0588 \\
A--LB & $(4H,4H,H)$ & $(512,512,128)$ & $100$ & $1.4 \times
10^{-2} \pm 1.4 \times 10^{-3}$ & 0.0945 & 0.0815 & 0.0573 \\
\hline\hline
\end{tabular}
\caption{Properties of MHD turbulence when forcing is turned
  off. The first column labels the model. The box size, resolution and
  duration (in orbits) of the simulations are given in columns two,
  three and four. The time averaged properties of the turbulence,
  i.e. the value of $\alpha$ (total stress normalized by the thermal pressure)
  and the velocity fluctuations along the three spatial (normalized by the
  speed of sound) appear in the following four columns.}
\label{table_noforcing}
\end{center}
\end{table*}

\begin{figure*}
\begin{center}
\includegraphics[scale=0.33]{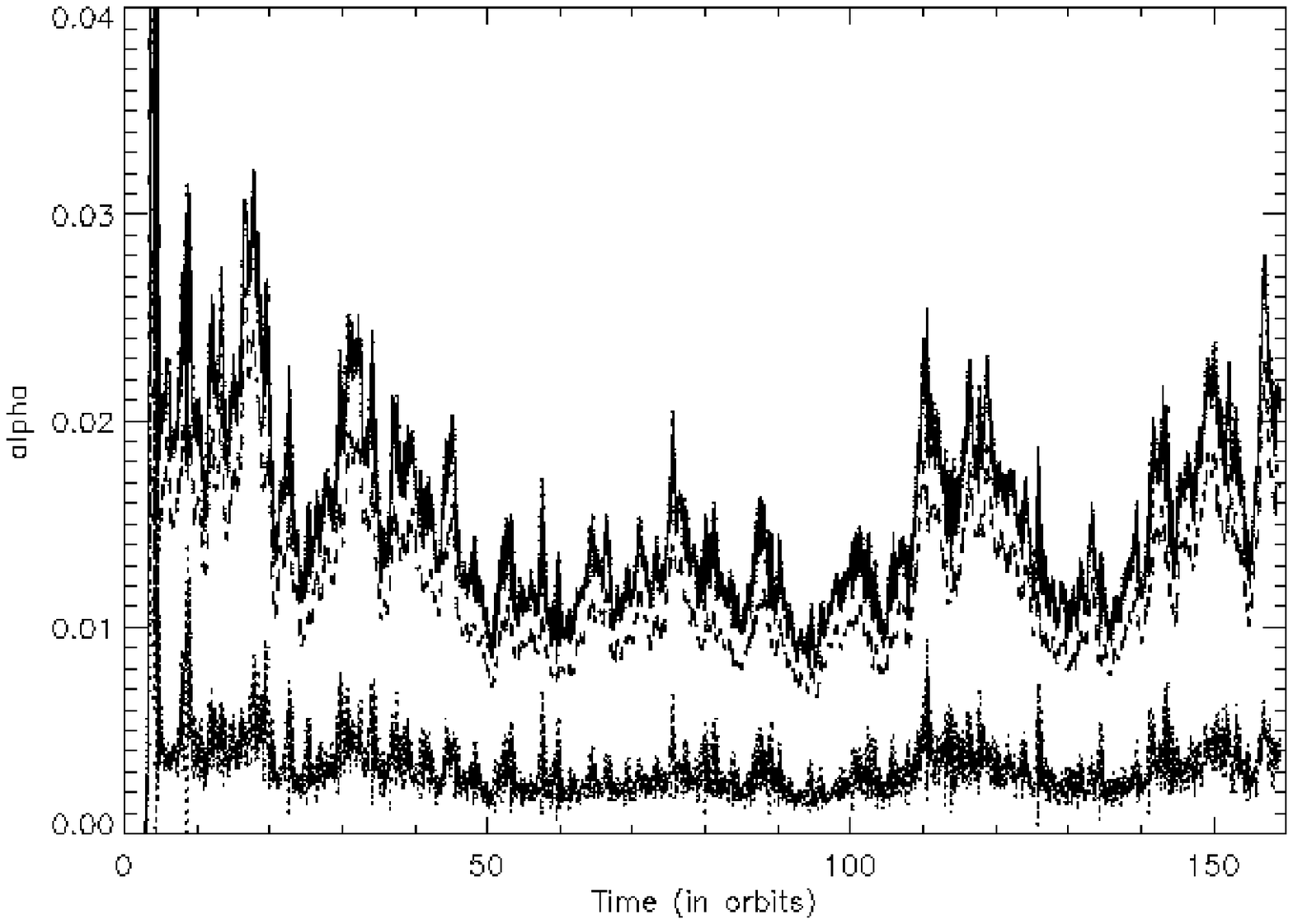}
\includegraphics[scale=0.33]{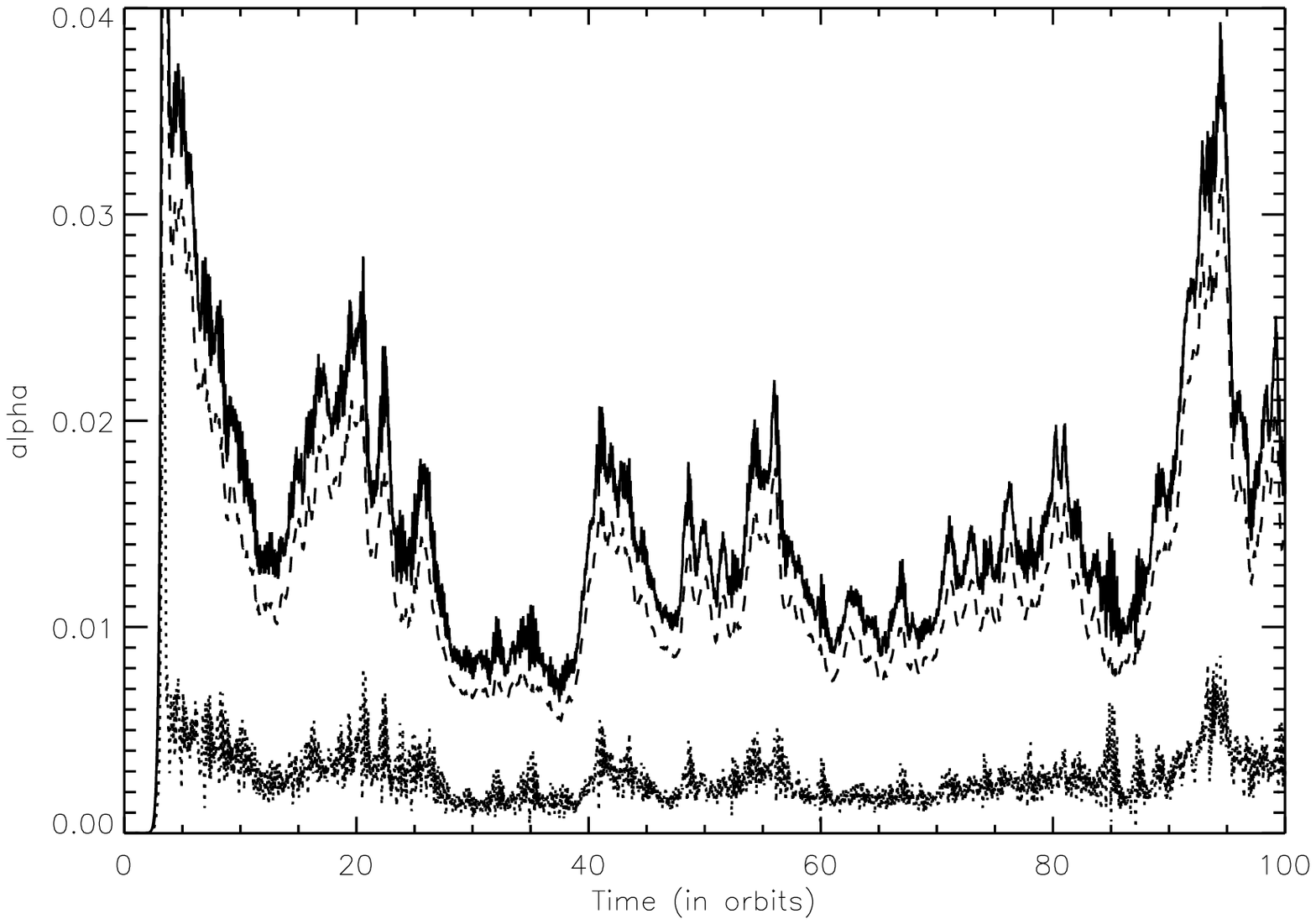}
\includegraphics[scale=0.33]{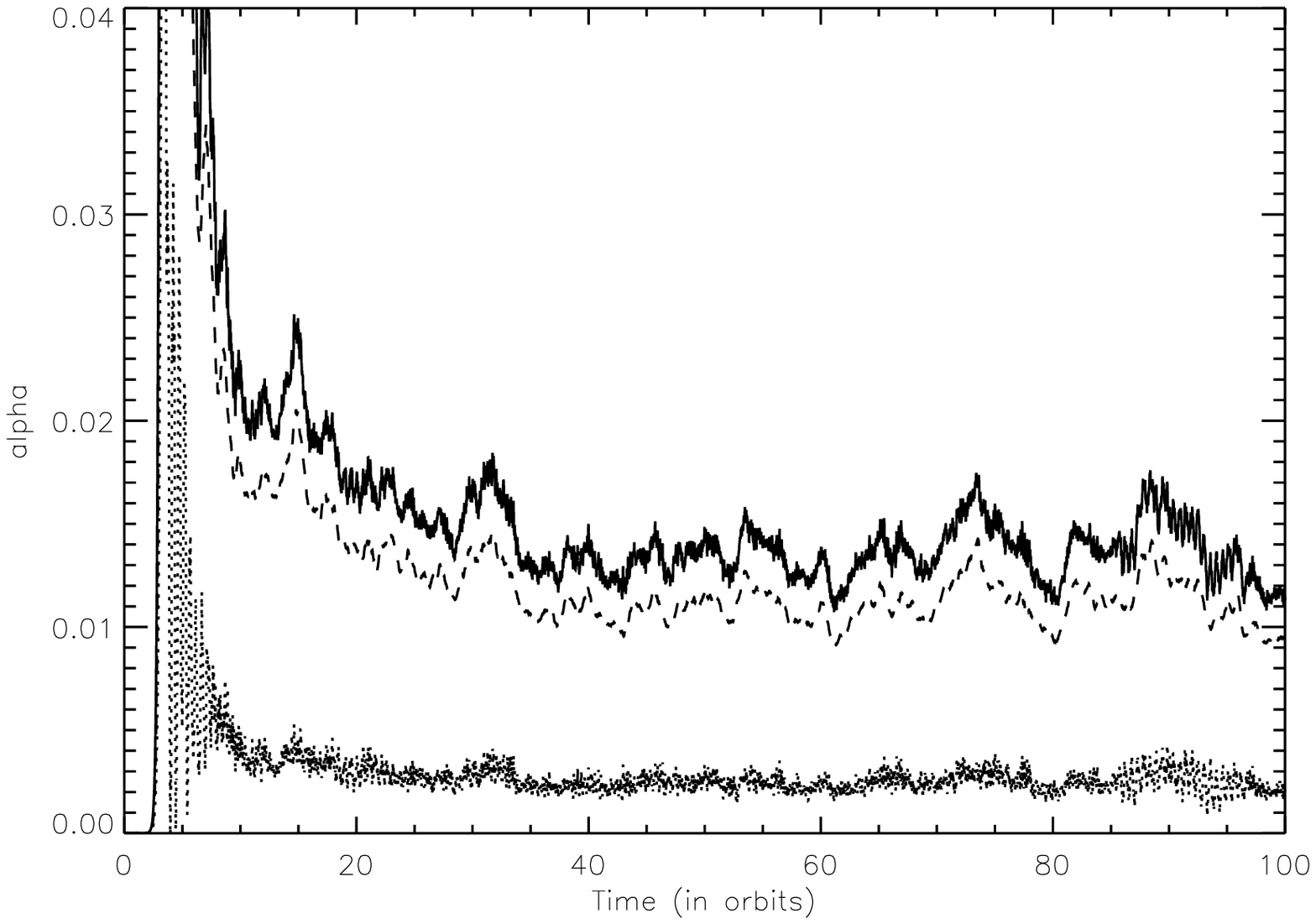}
\caption{Time history of $\alpha$ ({\it solid line}), the Maxwell
  stress $T_{r\phi}^{Max}$ ({\it dashed line}) and the Reynolds stress
  $T_{r\phi}^{Rey}$ ({\it dotted line}). From left to right, the three
  panels show the results of model {\it R--SB},
  {\it A--SB} and {\it A--LB} respectively. For all cases,
  $Re=3125$ and $Pm=4$.}
\label{alpha_code_compar}
\end{center}
\end{figure*}

\begin{figure*}
\begin{center}
\includegraphics[scale=0.25]{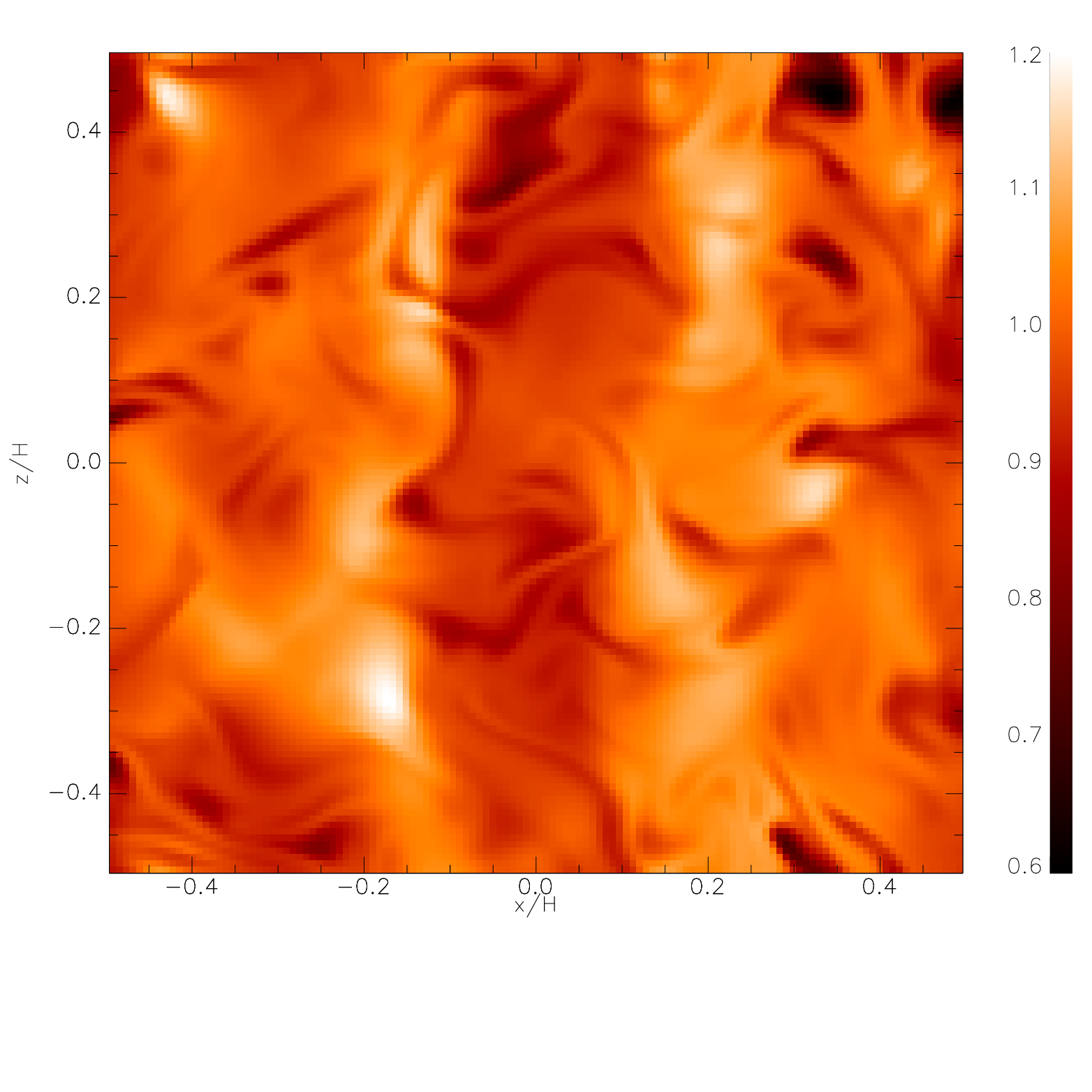}
\includegraphics[scale=0.25]{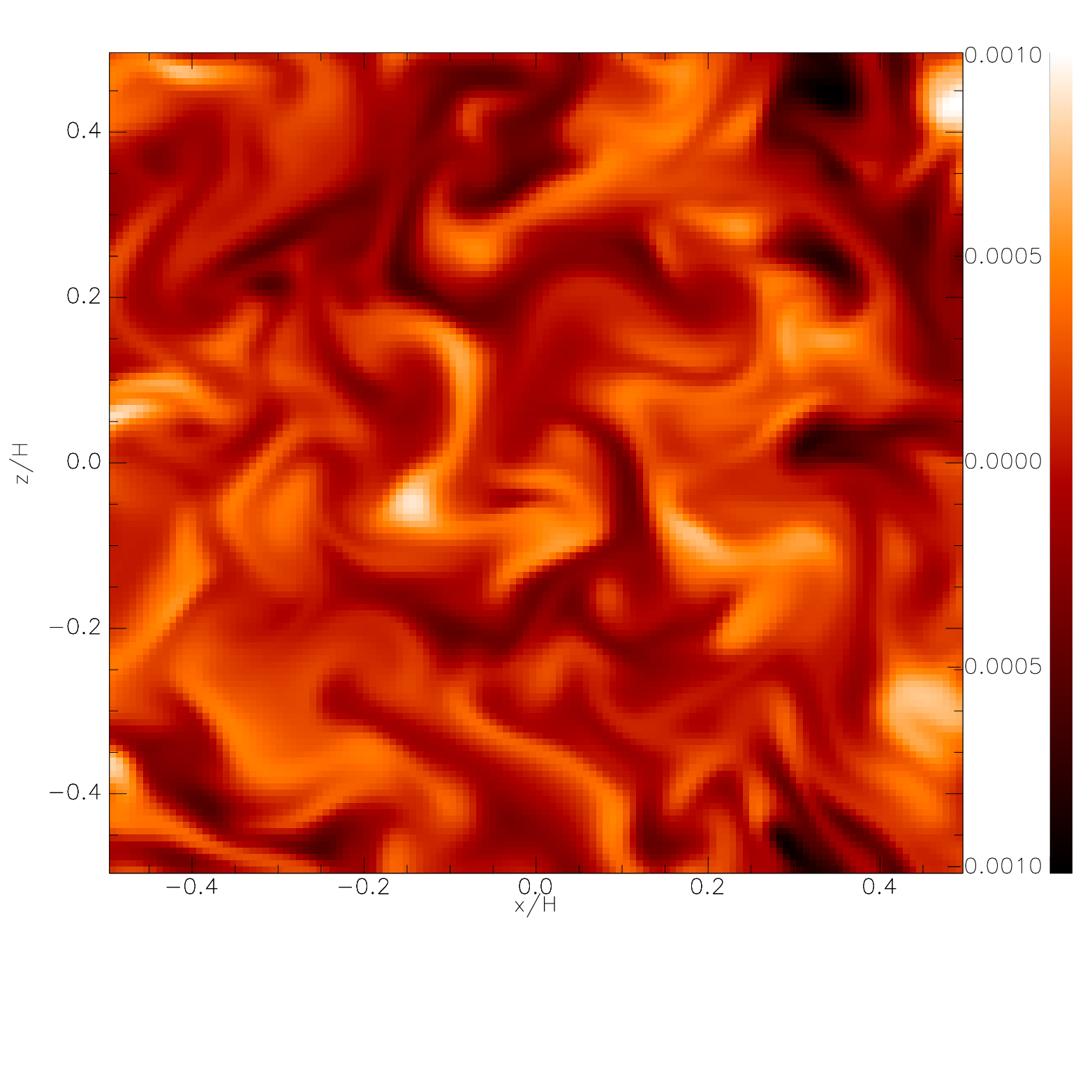}
\includegraphics[scale=0.25]{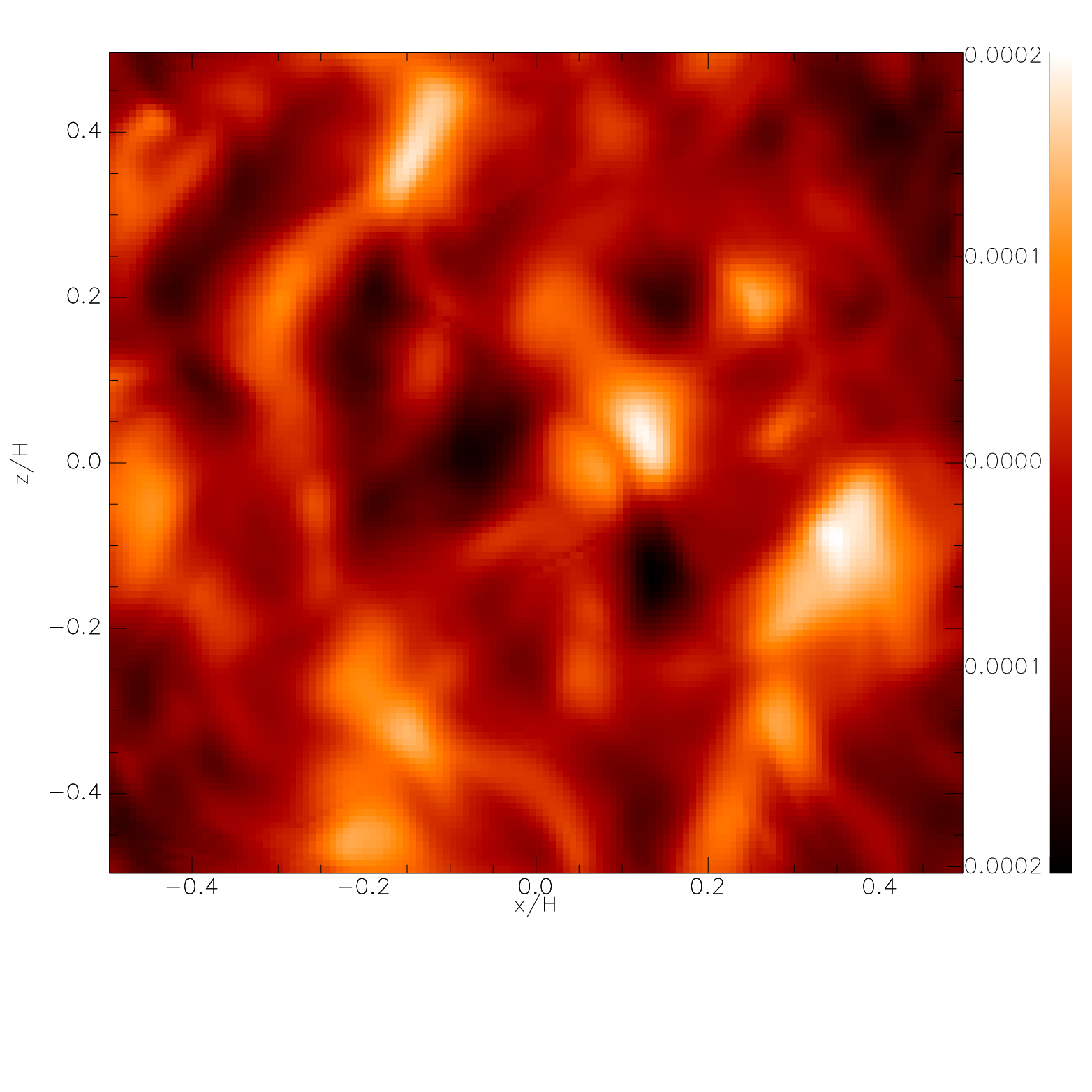}
\caption{Structure of the flow in the $(x,z)$ plane obtained in model
  {\it R--SB} at t=70 orbits. The left, middle and right panels
  show the density, azimuthal component of the magnetic field, and
  vertical component of the velocity respectively. Similar results are obtained
  in model {\it A--SB}.}
\label{snap_smallbox_dumses}
\end{center}
\end{figure*}

\begin{figure*}
\begin{center}
\includegraphics[scale=0.75]{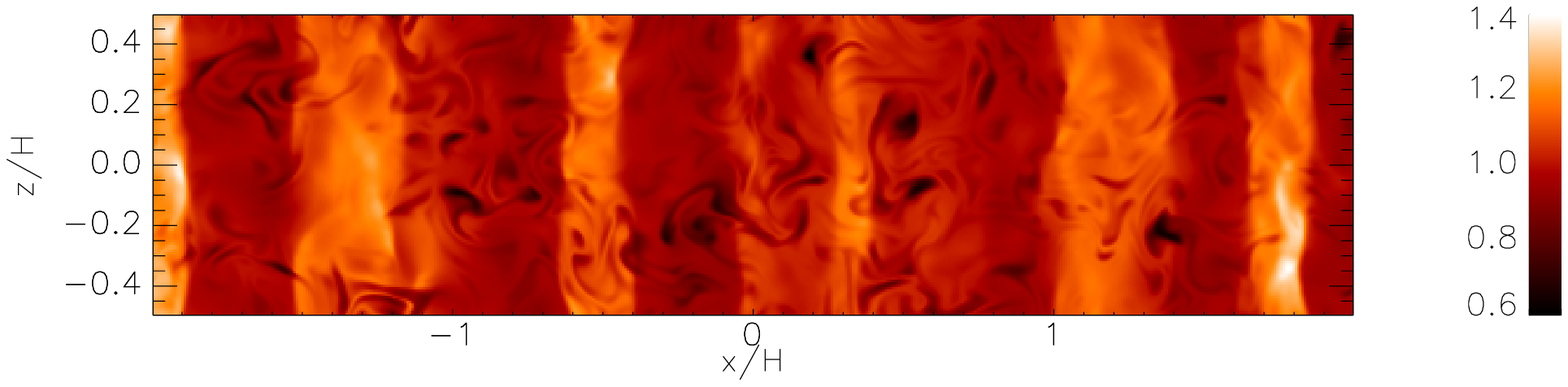}
\includegraphics[scale=0.75]{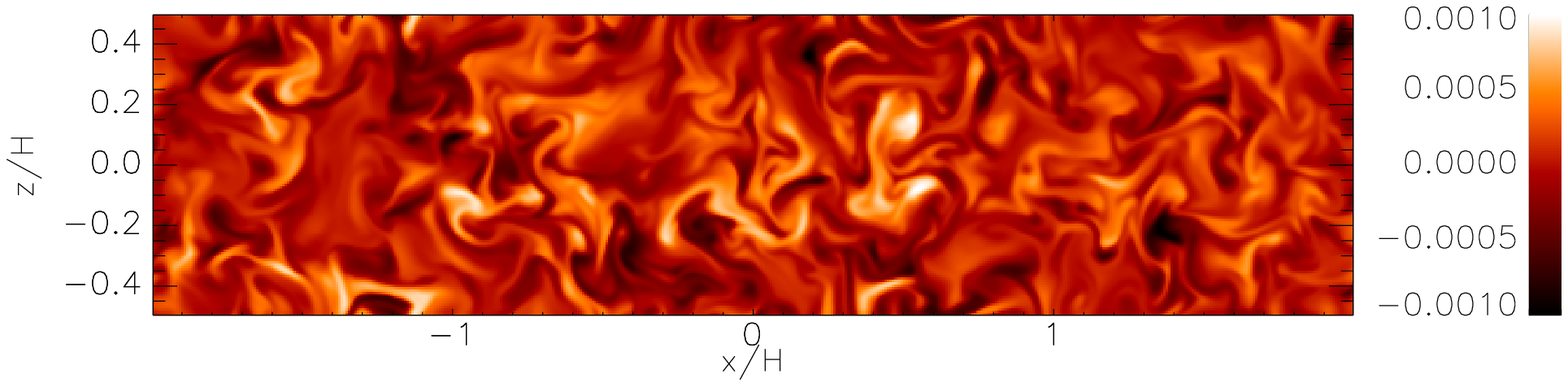}
\includegraphics[scale=0.75]{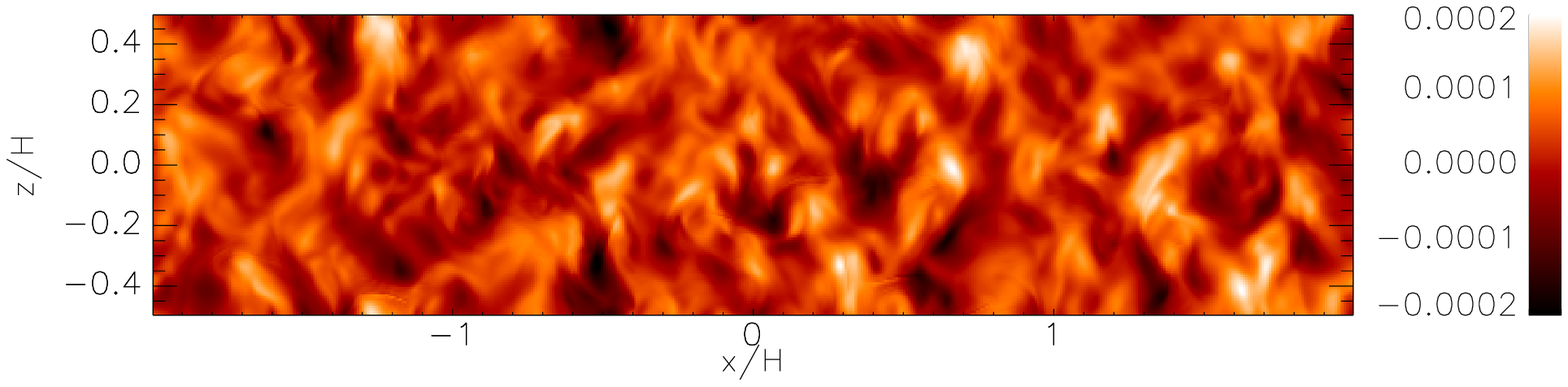}
\caption{Same as figure~\ref{snap_smallbox_dumses} but for model {\it A--LB}:
  the upper, middle and lower panels show the density,
  azimuthal component of the magnetic field, and vertical component of
  the velocity respectively.}
\label{snap_bigbox_ath}
\end{center}
\end{figure*}

\begin{figure*}
\begin{center}
\includegraphics[scale=0.75]{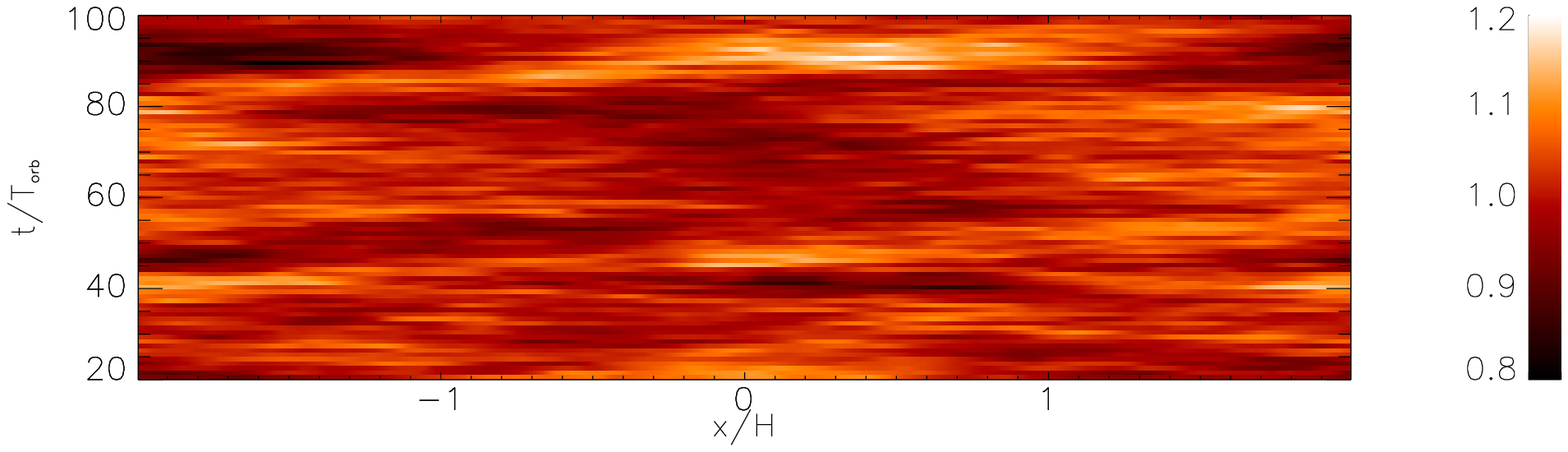}
\caption{Space--time diagram showing the radial variation of the
  density, averaged along y and z, in model {\it A--LB}. It
  shows the appearance of long--lived density features. They are
  associated with large scale zonal flows as recently reported by
  \citet{johansenetal09}.} 
\label{zonal_bigbox_ath}
\end{center}
\end{figure*}

The parameters of the runs we present in this section, along with
time averaged values from selected quantities measured from the runs, are
given in Table~\ref{table_noforcing}.  Each run is labeled according
to the code used (``R" for Ramses, or ``A" for Athena) and the
size of the box: ``SB'' for Small Box, i.e. simulations performed
with $(Lx,Ly,Lz)=(H,4H,H)$ and ``LB''
for Large Box, i.e. $(Lx,Ly,Lz)=(4H,4H,H)$. The resolution and 
duration of the runs are given in columns 3 and 4, respectively. In
column 5, we report for each model the value of $\alpha$ time averaged
between $t=50$ and the end of the simulation.  As is usual, it is defined
by the sum of the Reynolds and Maxwell stress tensors,
$T_{r\phi}^{Rey}$ and $T_{r\phi}^{Max}$, normalized by the volume
averaged thermal pressure $<P>$: 
\begin{equation}
\alpha=\frac{T_{r\phi}^{Rey}+T_{r\phi}^{Max}}{<P>}=\frac{\langle \rho
  (v_x-\bar{v_x})(v_y-\bar{v_y})\rangle-\langle B_x B_y \rangle}{<P>} \, ,
\end{equation}
where $\bar{v_x}$ and $\bar{v_y}$ are y and z averages of $v_x$ and
$v_y$ respectively, and $<>$ denotes a volume
average. Finally, the last three columns in
Table~\ref{table_noforcing} give the amplitude of the 
gas velocity fluctuations along the three spatial coordinates.

We performed three runs. Two of them, R--SB and A--SB, share identical
parameters but use different codes, Ramses and Athena respectively,
in order to compare the results from both codes. The third one, A--LB,
was performed with Athena in a larger domain. The last four columns
of Table~\ref{table_noforcing} show that the saturated state of the
turbulence is almost identical in the three models. For example,
$\alpha=0.014$, $0.015$ and $0.014$ respectively in model R--SB,
A--SB and A--LB. This agreement is confirmed by the three panels of
figure~\ref{alpha_code_compar} which correspond, from left to right, to
R--SB, A--SB and A--LB. All show the time history of $T_{r\phi}^{Rey}/<P>$
({\it lower dashed line}), $T_{r\phi}^{Max}/<P>$ ({\it upper dashed
line}) and $\alpha$ ({\it solid line}). The curves obtained in the three
simulations are very similar and confirm the time averaged $\alpha$ values
given in Table~\ref{table_noforcing}. These results have two implications.
First, the good agreement between R--SB and A--SB gives confidence in
both codes, as the two models have identical parameters and differ
only in the algorithms.  Thus, it validates both the implementation
of the shearing box model in Ramses and the implementation of orbital
advection in Athena.  Given the rms fluctuations in $\alpha$ reported
in Table~\ref{table_noforcing}, these results are also consistent with
the value of $\alpha \sim 0.01$ quoted by \citet{fromangetal07} for
their model {\it 128Re3125Pm4}. We note, nonetheless, that the results
obtained here with Athena and Ramses appear to lead to slightly larger
value of the angular momentum transport. It is unclear (and beyond the
scope of this paper) whether this difference is significant. It may
be partly due to our box size being slightly larger in the azimuthal
direction compared to that used in \citet{fromangetal07}, or it could
be due to small differences in the numerical dissipation between the
various codes.  The second result that emerges from the models presented
in this section is the insensitivity of the turbulence properties to the
box size.  The time averaged value of $\alpha$ in model A--LB is almost
identical to the other two. The time history presented on the right panel
of figure~\ref{alpha_code_compar} also shows good agreement with the
other two panels. The only difference is that the fluctuations around
the mean $\alpha$ value are smaller in the case of the big box. This
translates into a standard deviation which is about three times smaller
in A--LB than in the other two models. Again, the reasons for this
difference are unclear. It may be due to better statistics in the
large box (simply due to the larger volume of the simulations) that
average over extreme events. Alternatively, it might be
due to the larger number of parasitic modes than can develop in larger
boxes \citep{pessah&goodman09} or to a more efficient
nonlinear coupling between the more numerous turbulent modes present
in a larger box \citep{latteretal09} . Both effects would reduce the
lifetime 
(and therefore the influence) of channel modes. The similarity
between the small and large boxes model is further confirmed by
figure~\ref{snap_smallbox_dumses} and \ref{snap_bigbox_ath}. The
former shows snapshots of the density ({\it left panel}), $B_y$ ({\it
middle panel}) and $v_z$ ({\it right panel}) in the $(x,z)$ plane for
model R--SB at $t=70$ (very similar figures are obtained for model
A--SB). The equivalent figures for model A--LB at $t=70$ are shown in
figure~\ref{snap_bigbox_ath}.  The figure shows that the structure of
the flow in the large box is very similar to four patches of the small
box model repeated next to each other.

Other features typical of shearing box simulations of
the MRI, like density wave propagating radially in the box
\citep{heinemann&papaloizou09a,heinemann&papaloizou09b}, are clearly seen
in the density field regardless of the box size. Finally, as expected
for a simulation having $Pm>1$, the smallest scale structure seen in the
velocity field is generally larger than the smallest scale structure
seen in the magnetic field.  The latter shows elongated filaments
reminiscent of high $Pm$ small scale dynamo simulations \citep[see for
example][]{Schekochihin07b}.

Recently, \citet{johansenetal09} also reported numerical simulations
of MHD turbulence in the shearing box in large domains.  They found
an increase of $\alpha$ when going from small ($H$) to large ($4H$)
boxes. Their result differs from those reported here, where we find
no sensitivity of $\alpha$ on box size. The reason probably lies
in the fact that we always use domains with an azimuthal extend of
$4H$, even when the radial domain is small (i.e. when $L_x = H$).  On the
other hand, \citet{johansenetal09} used square domains, with $L_x =
L_y = H$ in their small box.  It is well known that the azimuthal
extent of the domain affects the saturated level of the turbulence
\citep{hawleyetal95,hawleyetal96}. We conclude that, for fixed
  and large azimuthal extent of the domain, the converged value of
$\alpha$ is insensitive to the radial dimension of the box, at least
for the values of the Reynolds and Prandtl numbers studied here.

Another property of the flow reported by \citet{johansenetal09} is
the existence of large scale, long--lived zonal flows that generate
axisymmetric density features in large boxes. We have looked for, and found,
such features in our large box simulations as well.  An example is
shown in figure~\ref{zonal_bigbox_ath}, a space--time
diagram of the density in model A--LB. It is similar to figure~6
of \citet{johansenetal09} and shows a similar amplitude and lifetime for
large scale density features, indicating that a zonal flows also develops
in our simulations. As both results were obtained using completely
different codes, our results confirm the conclusion drawn by
\citet{johansenetal09}: zonal flows appear to be a robust feature
in MRI-driven turbulence in the shearing box. 

This completes our description of the models computed in the absence
of a forcing EMF. The flow in these models provides a starting point
to study turbulent resistivity induced by the MRI.  Namely, at $t=30$,
we restarted the models using various amplitudes and directions for the
forcing EMF, $\bb{E}$. The following section describe these results.

\section{Turbulent resistivity}
\label{forcing_sec}

\begin{table*}[t]\begin{center}\begin{tabular}{@{}lccccccccc}\hline\hline
Model & $t_{avg}$ & $|\bb{E}|$ & $\alpha$ & $\delta v_x/c_0$ & $\delta v_y/c_0$
& $\delta v_z/c_0$ & $\beta_{max}$ & $\eta_t$ & $Pm_t$ \\
\hline\hline
Ey-4E-10-R & 50 & $4 \times 10^{-10}$ & $2.3 \times 10^{-2}$ &
0.1214 & 0.1060 & 0.0789 & $5.6 \times
10^{4}$ & $1.07 \times 10^{-5}$ & $1.47$  \\
Ey-4E-10-R-LONG & 120 & $4 \times 10^{-10}$ & $2.0 \times 10^{-2}$ &
0.1173 & 0.1001 & 0.0758 & $5.5 \times
10^{4}$ & $1.06 \times 10^{-5}$ & $1.27$  \\
Ey-1E-9-R & 50 & $1 \times 10^{-9}$ & $2.6 \times 10^{-2}$ &
0.1344 & 0.1231 & 0.0896 & $1.7 \times
10^{4}$ & $1.46 \times 10^{-5}$ & $1.21$  \\
Ey-4E-9-R & 50 & $4 \times 10^{-9}$ & $5.2 \times 10^{-2}$ &
0.1831 & 0.1885 & 0.1246 & $1.6 \times
10^{3}$ & $1.81 \times 10^{-5}$ & $1.91$  \\
Ey-8E-9-R & 50 & $8 \times 10^{-9}$ & $7.6 \times 10^{-2}$ &
0.2209 & 0.2458 & 0.1493 & $7.3 \times
10^{2}$ & $2.43 \times 10^{-5}$ & $2.10$  \\
\hline
Ey-4E-10-A & 50 & $4 \times 10^{-10}$ & $1.7 \times 10^{-2}$ &
0.1020 & 0.0846 & 0.0633 & $5.8 \times
10^{4}$ & $1.08 \times 10^{-5}$ & $1.02$  \\
Ey-4E-9-A & 50 & $4 \times 10^{-9}$ & $5.3 \times 10^{-2}$ &
0.1766 & 0.1876 & 0.1189 & $1.2 \times
10^{3}$ & $1.56 \times 10^{-5}$ & $2.28$  \\
\hline
Ey-4E-10-A-L4-lx4 & 48 & $4 \times 10^{-10}$ & $2.6 \times 10^{-2}$ &
0.1242 & 0.1202 & 0.0798 & $2.8 \times
10^{4}$ & $3.02 \times 10^{-5}$ & $0.57$  \\
Ey-4E-9-A-L4-lx4 & 50 & $4 \times 10^{-9}$ & $1.0 \times 10^{-1}$ &
0.2371 & 0.2624 & 0.1555 & $1.4 \times
10^{3}$ & $6.68 \times 10^{-5}$ & $1.00$  \\
Ey-4E-9-A-L4-lx1 & 50 & $4 \times 10^{-9}$ & $5.3 \times
10^{-2}$ & 0.1789 & 0.1875 & 0.1204 & $1.6 \times 10^{3}$ & $1.82 \times 10^{-5}$ & $1.93$  \\
\hline
Ex-4E-9-R & 50 & $4 \times 10^{-9}$ & $2.4 \times 10^{-2}$ &
0.1292 & 0.1162 & 0.08476 & $5.3 \times
10^{2}$ & $1.05 \times 10^{-5}$ & $1.57$  \\
Ex-8E-9-R & 50 & $8 \times 10^{-9}$ & $4.6 \times 10^{-2}$ &
0.1746 & 0.1747 & 0.1160 & $1.3 \times
10^{2}$ & $1.01 \times 10^{-5}$ & $3.00$  \\
\hline\hline
\end{tabular}
\caption{Properties of MHD turbulence when a forcing EMF is turned on. The
first column gives the model label, the second the duration over which the data
are averaged, and the third the amplitude of the forcing EMF
used in that case.  Subsequent columns give time averaged values of
various quantities of physical interest, namely the mean values of
$\alpha$, the maximum value of $\beta$ (ratio between the thermal and
magnetic pressure), the value of the turbulent resistivity that
results from fitting the mean vertical magnetic field, and the
turbulent magnetic Prandtl number of the flow.}
\label{table_forcing}
\end{center}
\end{table*}

The properties and results of the runs we performed are listed
in Table~\ref{table_forcing}. The first column gives a label
for the name, coded according to the following convention:
``direction--amplitude--code--parameters'' where ``direction'' can be
either ``Ey'' or ``Ex'' and indicates the direction of the forcing EMF
(we considered only forcing aligned along either the radial or the
azimuthal direction, respectively creating vertical or toroidal field),
``amplitude'' gives the amplitude of $\bb{E}$, ``code'' reports the code
we used for that model (``R" for Ramses, and ``A" for Athena), while
``parameters'', whenever present, gives additional parameters that will
be specified when needed in the text. For example, model {\it Ey-4E-10-R}
was performed with Ramses using a forcing EMF in the azimuthal direction
of amplitude $4 \times 10^{-10}$. The other parameters reported in
Table~\ref{table_forcing} are the duration over which the data were
averaged (column 2), the amplitude of the forcing EMF (column 3),
the value of $\alpha$ (column 4) and the amplitude of the velocity
fluctuations along the three spatial coordinates (column 5, 6 and 7).

The forcing EMF we imposed on the flow results in a steady-state magnetic
field profile. We fitted the profile (averaged in time over the
duration of the run and in space over the two directions perpendicular
to the direction of variation) by a sinusoidal
function. The amplitude of the fit, $B_{fit}$, is expressed via the parameter
\begin{equation}
\beta_{max}=\frac{<P>}{B_{fit}^2/2} \, ,
\end{equation}
and given in column 8 on Table~\ref{table_forcing}. The turbulent
resistivity $\eta_t$ responsible for that amplitude is calculated using
Eq.~(\ref{eta_turb_x}) or (\ref{eta_turb_z}), depending of the
direction of the forcing EMF. Its value is reported in column 9.

Finally, we follow \citet{guan&gammie09} and associate a turbulent
``viscosity'' $\nu_t$ with the flow, where
\begin{equation}
\nu_t=\frac{2}{3}\alpha c_0 H,
\end{equation}
which we use to define a turbulent magnetic Prandtl number
$Pm_t=\nu_t/\eta_t$ given in column 10.

\subsection{Radial diffusion of a vertical magnetic field}

\subsubsection{$E_{0y}=4 \times 10^{-10}$}

\begin{figure*}
\begin{center}
\includegraphics[scale=0.4]{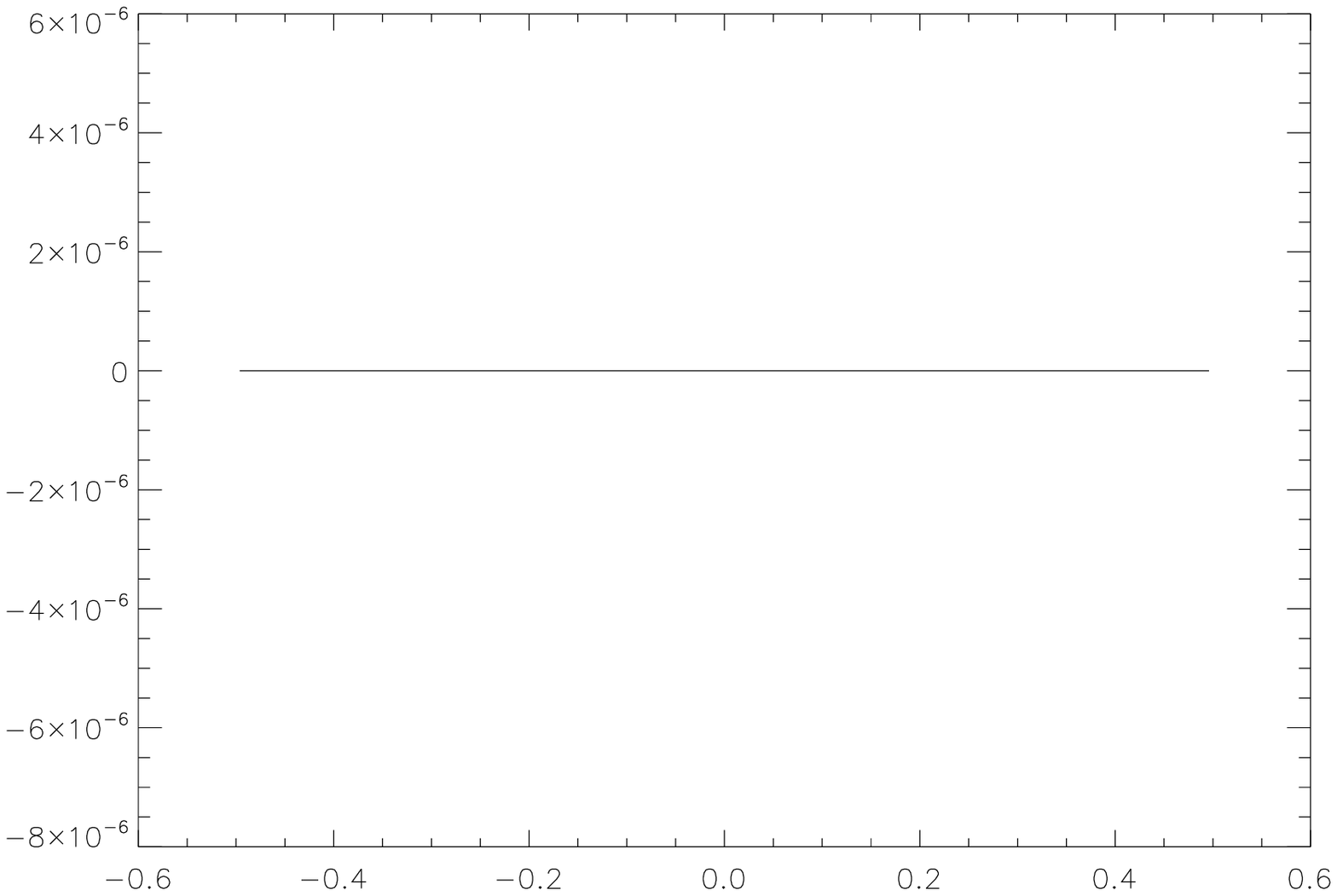}
\includegraphics[scale=0.4]{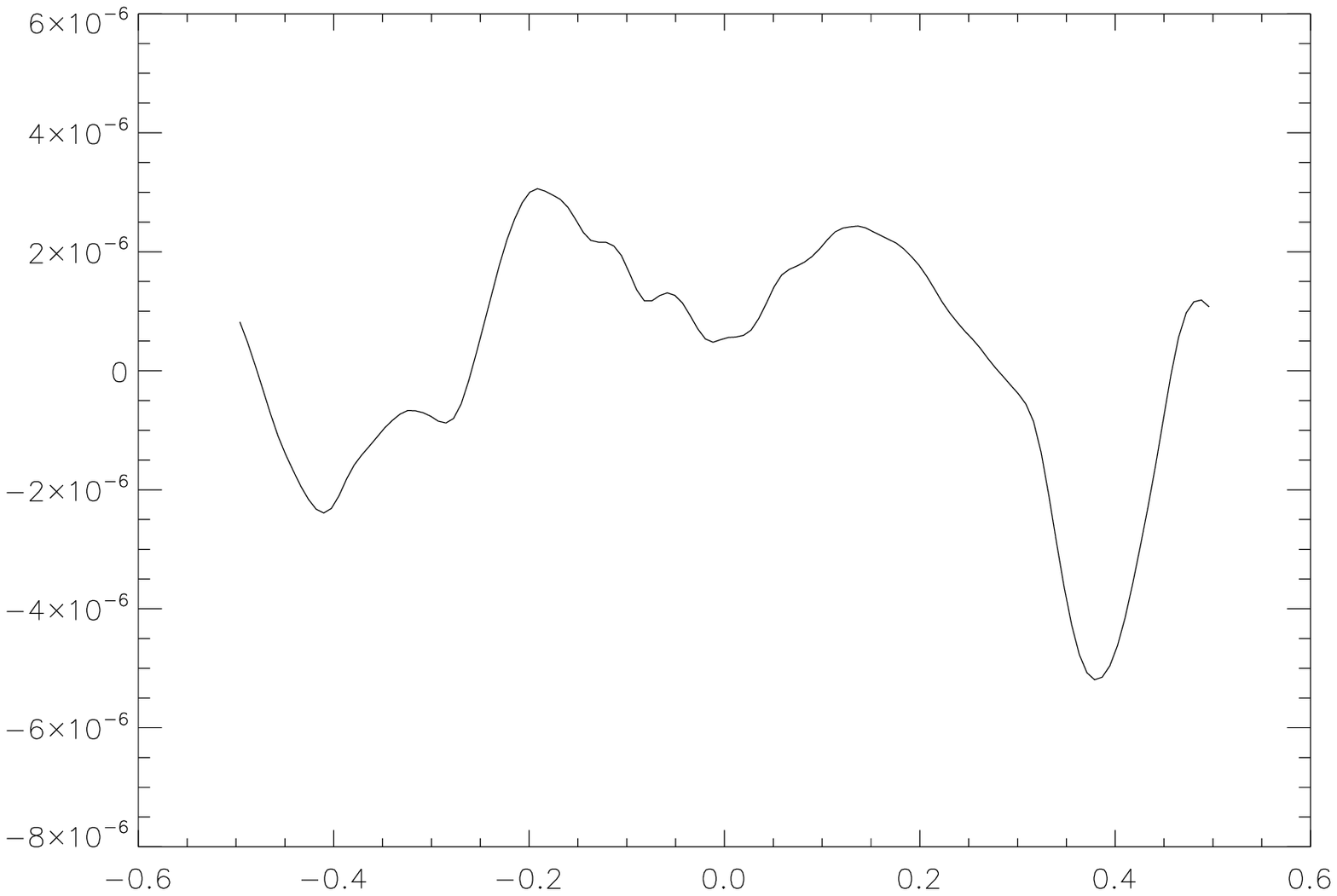}
\includegraphics[scale=0.4]{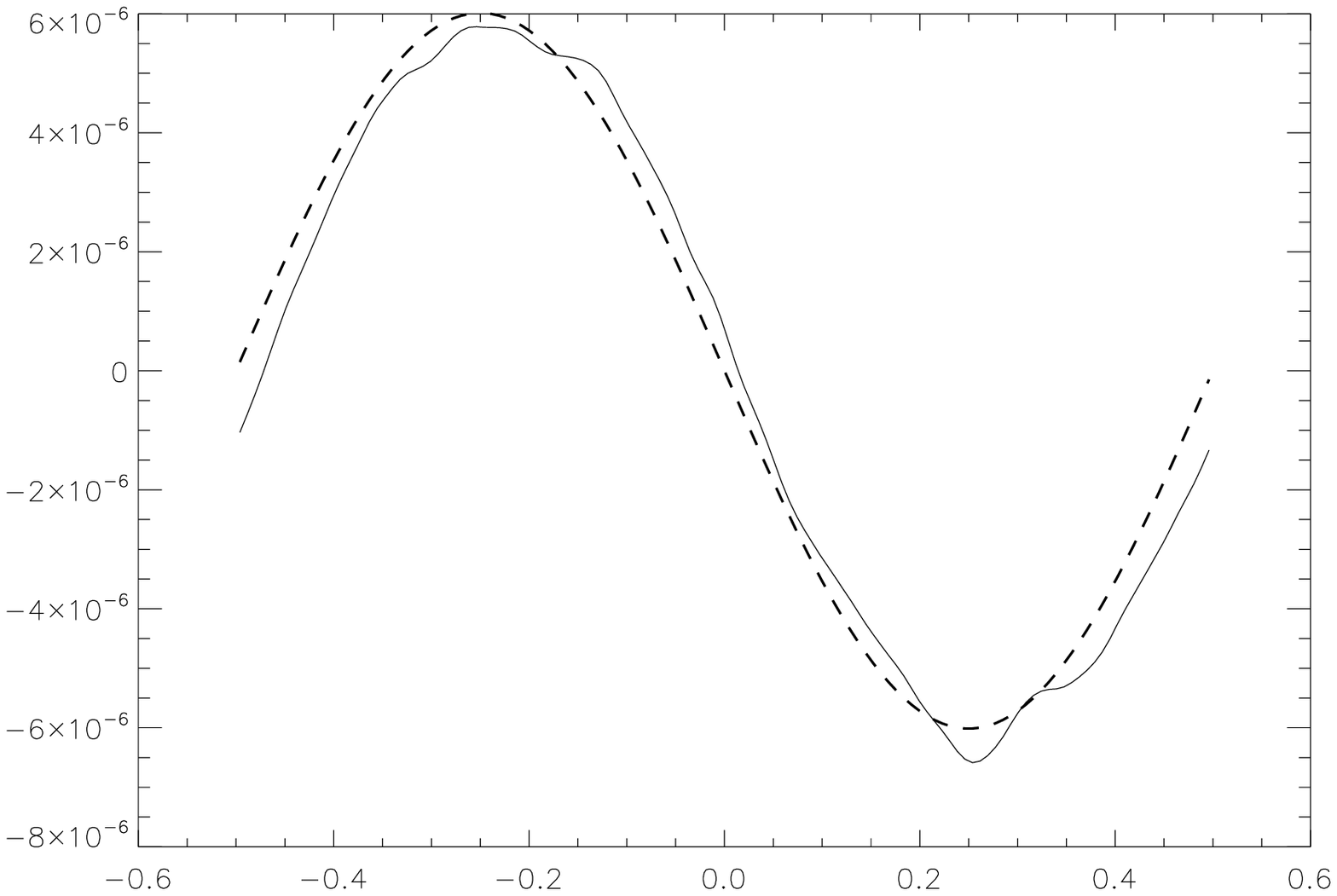}
\includegraphics[scale=0.4]{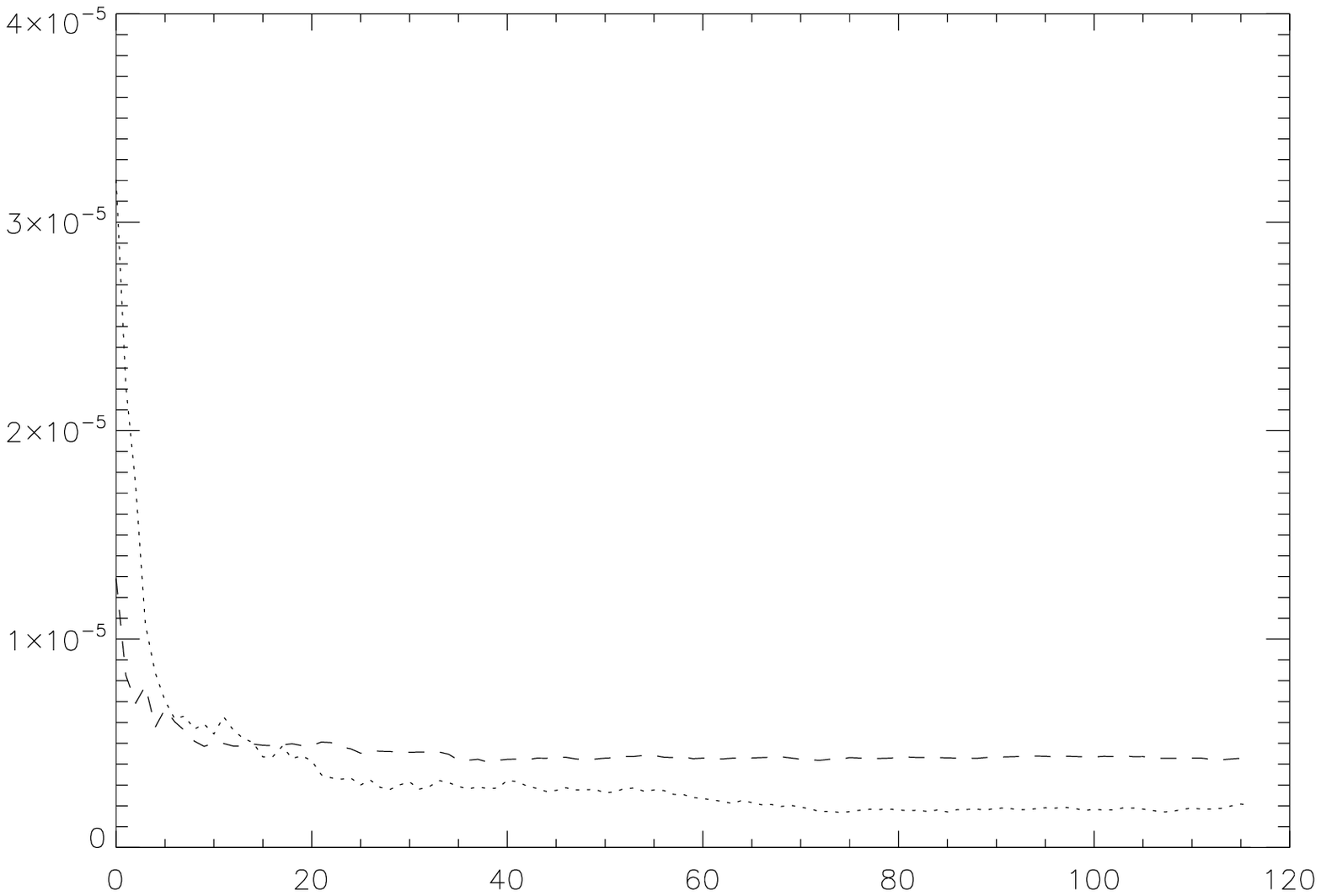}
\caption{Time averaged radial profile of the vertically and azimuthally
  averaged radial ({\it upper left panel}), azimuthal ({\it upper right
    panel}) and vertical ({\it lower left panel}) magnetic field for
  model {\it Ey-4E-10-R-LONG}. On the third panel, the solid line
  shows the curve extracted from the numerical simulations while the
  dashed line is a least-squared fit to the data assuming a sinusoidal
  profile. The lower right panel shows the time variation of the
  standard deviation of the three magnetic field components,
  plotted using solid, dotted and dashed lines respectively. These plots
  illustrate that the forcing EMF creates, as expected, a
radially varying vertical field with an amplitude larger than
the fluctuations of the other magnetic field components}
\label{e0_4.e-10_long}
\end{center}
\end{figure*}

In this section, we describe in detail the results we obtained for
models {\it Ey-4E-10-R}, {\it Ey-4E-10-R-LONG} and {\it Ey-4E-10-A}.
All were obtained using a forcing EMF directed along the
azimuthal direction, and varying only in the radial direction: $E_{0x}=0$
and $E_{0y}=4 \times 10^{-10}$. We used $\lambda_{E,x}=H$. The first
two were computed using 
Ramses and differ only in the duration of the averaging procedure we
applied. The third was computed using Athena and serves both to
validate our method and to quantify the uncertainty in our estimate of
the turbulent resistivity and magnetic Prandtl number.

The results we obtained for this set of parameters are
illustrated on figure~\ref{e0_4.e-10_long} for model {\it Ey-4E-10-R-LONG}.
There are four panels.  The solid line in
the first three (upper left, upper right and lower left)
plots the radial profile of $B_x$, $B_y$ and $B_z$ respectively,
averaged in time between $t=30$ and $t=150$ using $120$ dumps, and
averaged in space over $y$ and $z$. All plots use the same vertical
scale. It is apparent from the lower left panel that the toroidal forcing EMF
created a sinusoidally varying vertical field, as expected
from the discussion in section~\ref{setup}. The dashed line shows the
sinusoidal curve obtained by a least square fit to the data, and
gives an amplitude $B_{0z} \sim 6 \times 10^{-6}$ (this translates
into a value $\beta_{max}=5.5 \times 10^4$ as shown in
Table~\ref{table_forcing}). It is immediately obvious than this rather
low value is mostly due to the effect of turbulence. Indeed, using
Eq.~(\ref{steady_b}), the effect of the microscopic resistivity alone
would give an amplitude $B_{0z}^{\eta} \sim 8 \times 10^{-4}$,
larger by about two orders of magnitude than the measured value of
the vertical field. Using Eq.~(\ref{eta_turb_x}), we converted
$B_{0z}$ into a turbulent resistivity $\eta_{turb,x}=1.06 \times
10^{-5}$. Combined with the time averaged value of $\alpha=2.0 \times
10^{-3}$, this gives a turbulent magnetic Prandtl number of
$Pm_t=1.27$.

Given the rather low value of the forcing EMF, it is reasonable
to question the accuracy of these measurements. Indeed, the second
and third panels in figure~\ref{e0_4.e-10_long} indicate that the
fluctuations in $B_y$ ({\it upper right panel}) are of comparable
amplitude to $B_{0z}$\footnote{The absence of fluctuations in the $y$
and $z$ averaged radial profile of $B_x$ comes from the solenoidal
nature of the magnetic field}. The fourth ({\it lower right}) panel,
which shows the time history of the rms fluctuations in the running time
and spatial averages of $B_z$ ({\it solid line}) and $B_y$ ({\it dashed
line}), indicates a similar trend: although the rms fluctuations of $B_z$
become larger than those of $B_y$ after about $20$ orbits (thus indicating
than the standard averaging duration of $50$ orbits we usually take in
the following is enough), the ratio between the two is only a factor of
about two after $120$ orbits. In other words, the vertical field generated
by the forcing EMF is not much larger than the turbulent fluctuations in
the field. As such, our measurements are susceptible to a rather large
uncertainty.\footnote{The reason to study such small fields is so that
they do not significantly affect the property of the turbulence itself
(the wavelength of the most unstable wavelength of the MRI associated
with the field generated by the forcing EMF is largely underesolved in our
simulations).  Thus, they merely act as a passive probe of the turbulence,
which is not the case for the larger amplitude of $\bb{E}$ we consider
in section~\ref{forcing_var}}  In order to quantify this uncertainty,
we consider model {\it  Ey-4E-10-R}, which is the same as model {\it
Ey-4E-10-R-LONG} but averaged over only $50$ orbits, and model {\it
Ey-4E-10-A}, identical to model {\it  Ey-4E-10-R} but performed using
Athena. As shown on Table~\ref{table_forcing}, we found extremely close
values for the turbulent resistivities for all three models. The measured
turbulent magnetic Prandtl number varies by about $50 \%$, from $Pm_t \sim
1$ to $Pm_t \sim 1.5$. This difference can immediately be attributed to
variations in the measured $\alpha$, ranging from $1.7 \times 10^{-3}$
to $2.3 \times 10^{-3}$. Such a variation is compatible with the standard
deviation of $\alpha$ reported in Table~\ref{table_noforcing} and we
conclude therefore that a safe estimate of the turbulent Prandtl number
in the radial direction in this case lies in the range $[1.,1.5]$.

This estimate is in rough agreement with the recent results reported
by \citet{guan&gammie09} and \citet{lesur&longaretti09} in simulations
performed in the presence of a net magnetic field in the azimuthal
or vertical direction. It is also reassuring that our measured value
for the turbulent resistivity is in rough agreement with
mean field theories like the Second-Order Correlation Approximation
\citep[SOCA,][]{radler&rheinhardt07} also known as the First Order
Smoothing Approximation \citep[FOSA,][]{brandenburg&subramanian05}. In
this approach, the turbulent resistivity is expected to take the form
\begin{equation}
\eta_{t,x}^{SOCA}=\delta v_x^2 \tau_{corr} \, ,
\label{soca_eq}
\end{equation}
where $\tau_{corr}$ is the correlation time of the turbulent velocity
fluctuations. Estimates for $\tau_{corr}$ are not straightforward
to obtain. In a set of local and global stratified simulations of
MRI--induced turbulence having $\alpha \sim 0.01$, \citet{fromang&pap06}
and \citet{fromang&nelson09} found $\tau_{corr} \sim 0.15 T_{orb}$.
Adopting this estimate here, and using $\delta v_x \sim 0.11 c_0$ given
in table~\ref{table_forcing} for the three models we consider in this
section, Eq.~(\ref{soca_eq}) then gives
\begin{equation}
\eta_{t,x}^{SOCA}=1.14 \times 10^{-5} \, ,
\end{equation}
a value that is, given the approximations involved in the SOCA, in very
good agreement with the results reported in Table~\ref{table_forcing}.

\subsubsection{Varying the forcing EMFs}
\label{forcing_var}

\begin{figure*}
\begin{center}
\includegraphics[scale=0.33]{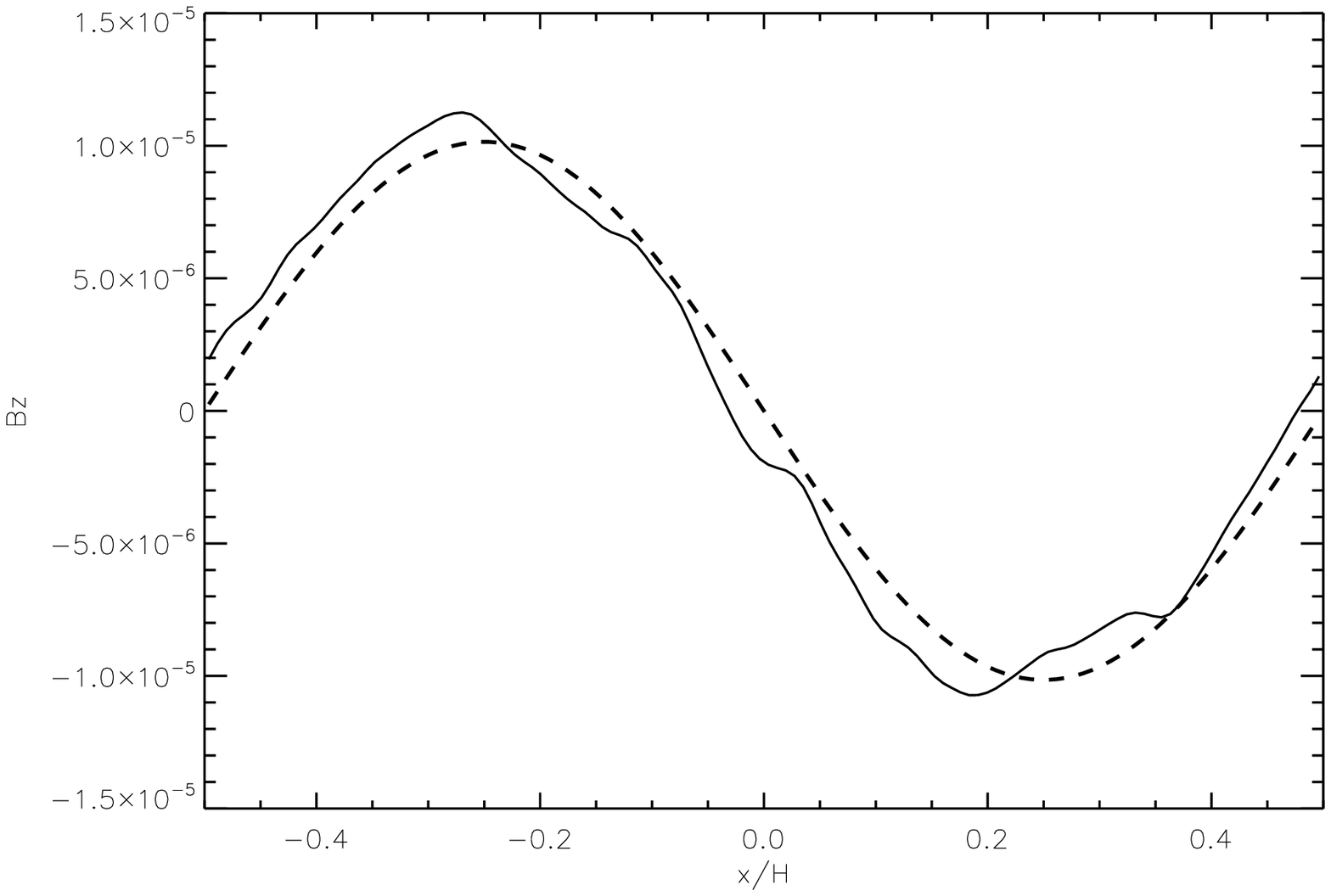}
\includegraphics[scale=0.33]{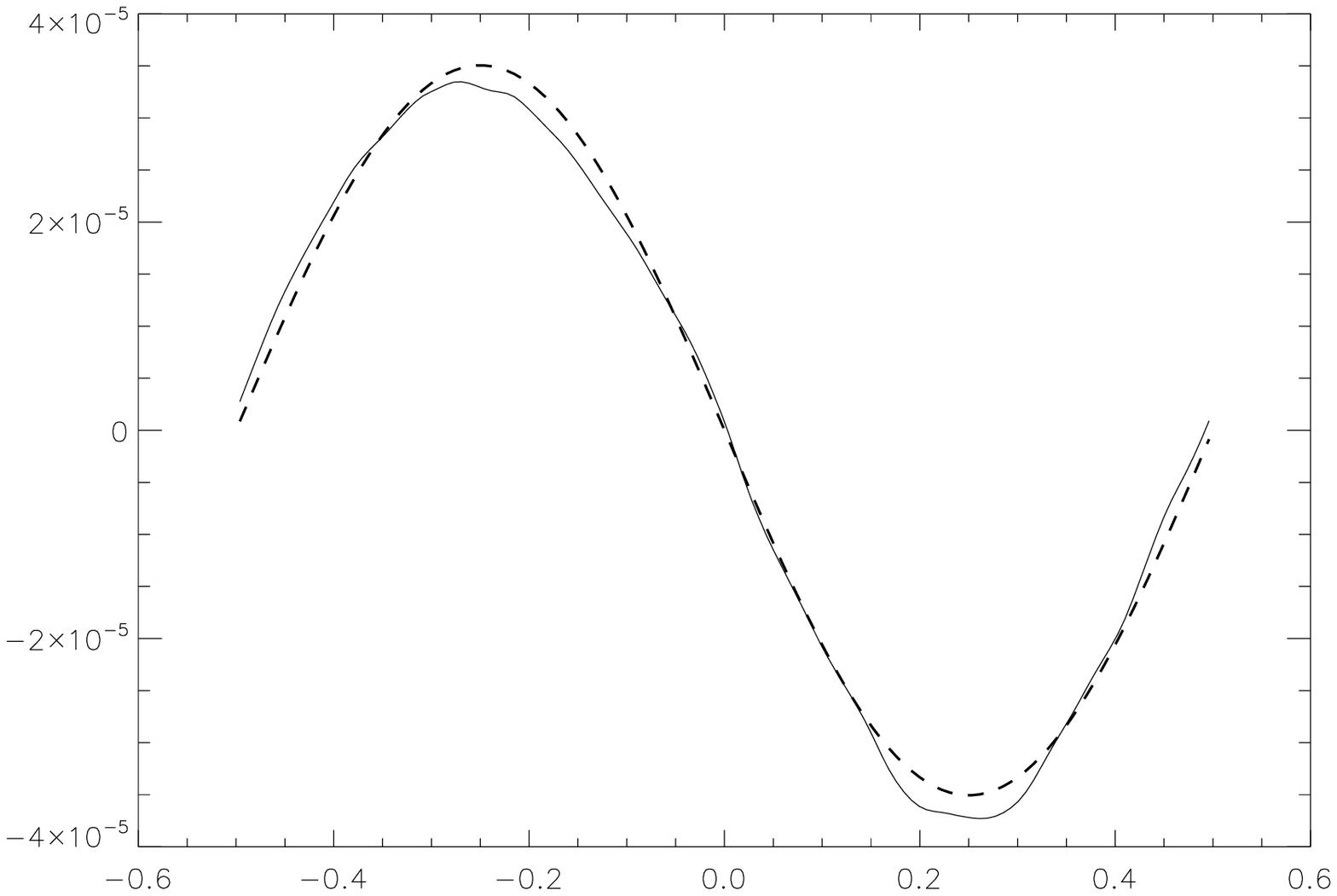}
\includegraphics[scale=0.33]{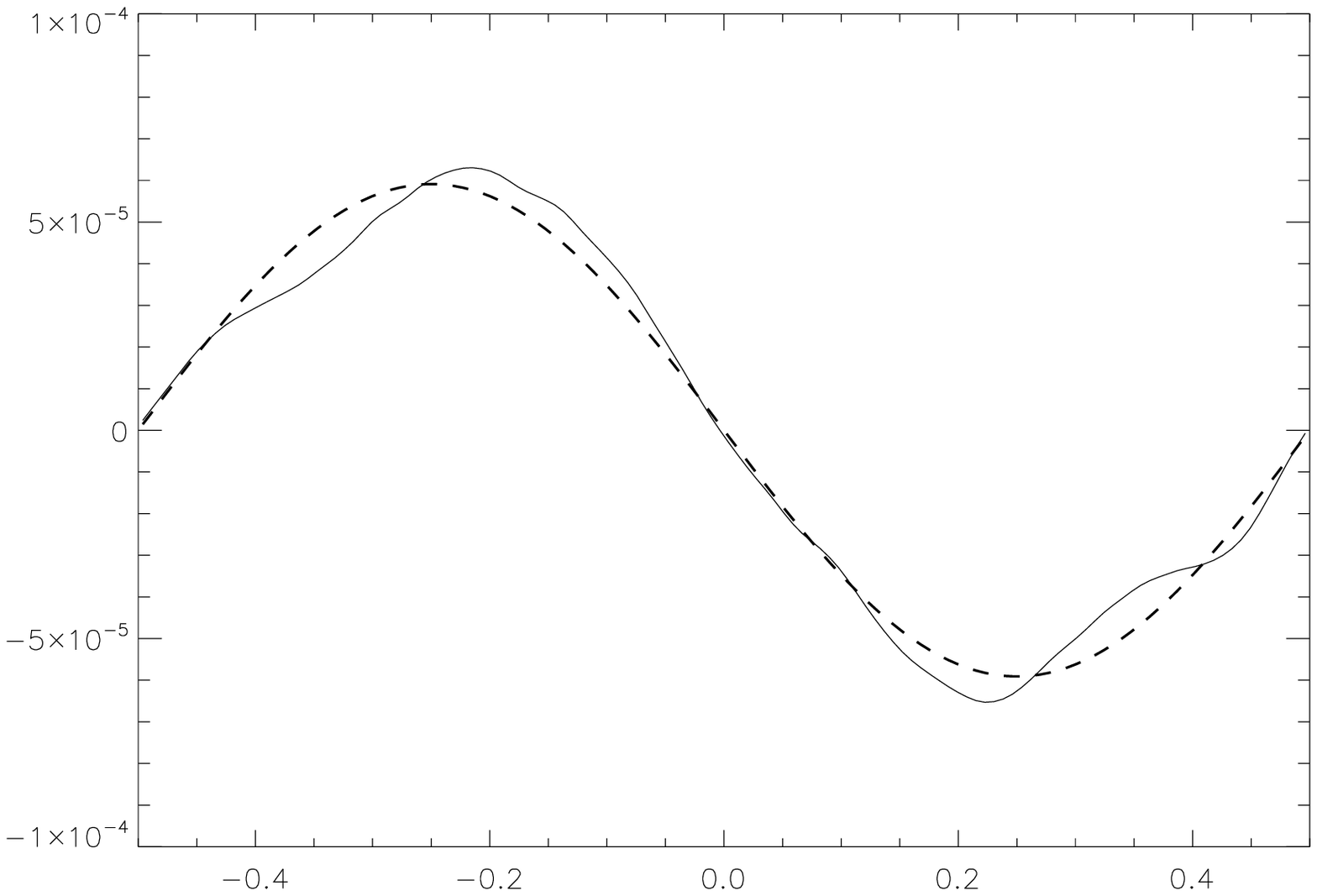}
\caption{Vertically, azimuthally and time averaged radial profile of
  the vertical magnetic field for models {\it Ey-1E-9-R}
  ({\it left panel}), {\it Ey-4E-9-R} ({\it middle panel}) and
  {\it Ey-8E-9-R} ({\it right panel}). On each panel, the dashed line
  is a sinusoidal fit to the simulation data used to estimate the
  turbulent resistivities reported in Table~\ref{table_forcing}.}
\label{mean_bz_vs_x}
\end{center}
\end{figure*}

\begin{figure}
\begin{center}
\includegraphics[scale=0.5]{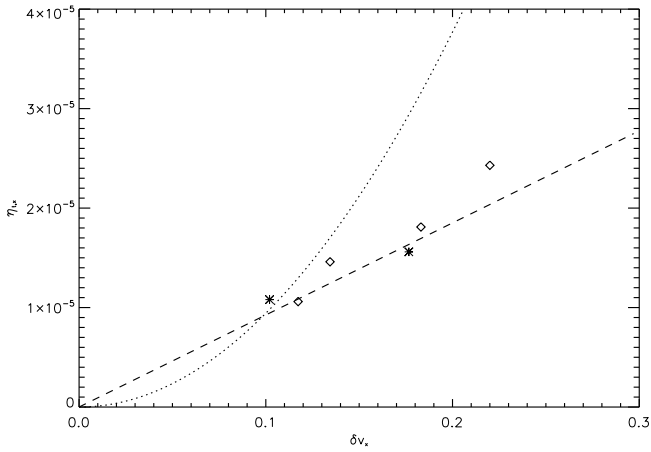}
\caption{Turbulent resistivity in the radial direction as a function
  of the turbulent velocity fluctuations for various models computed
  using different amplitude of the forcing EMFs. Diamonds correspond
  to run performed using Ramses while stars use the results of models
  computed with Athena. The dotted line shows the prediction of SOCA
  assuming a 
constant correlation time, independent of the velocity fluctuations,
while the dashed line shows the same prediction but computed assuming
a correlation lengthscale for the magnetic field independent of the
amplitude of the velocity fluctuations ({\it see text for
  details}). The latter is seen to be in good agreement with the data.}
\label{eta_x_plot}
\end{center}
\end{figure}

In the previous section, we considered a forcing whose amplitude only 
produces a very weak effect on the underlying turbulence. In this
section, we relax this hypothesis by considering larger forcing
EMFs. Such forcing may have an effect on the saturated properties of
the turbulence itself, which in turn may change the values of the
turbulent resistivity and $Pm_t$.

In the run we performed, listed in Table~\ref{table_forcing}, the
amplitude of the EMF was varied between $|\bb{E}|=4 \times 10^{-10}$
(described in the previous section) and $|\bb{E}|=8 \times 10^{-9}$,
keeping $\lambda_{E,x}=H$ in all cases. We tried larger values as well
but found that, regardless of the code we used, the flow developed
regions of low density and large magnetic field.  In these regions
  the large Alfven speed resulted in an extremely small timestep, which
  prevented the simulations from being continued.

Figure~\ref{mean_bz_vs_x} shows the radial profile (time averaged for
$50$ orbits and spatially averaged over the azimuthal and vertical
directions) of the vertical magnetic field we obtained in models {\it
Ey-1E-9-R} ({\it left panel}), {\it Ey-4E-9-R} ({\it middle panel}) and
{\it Ey-8E-9-R} ({\it right panel}). As was the case for the lower left
panel of figure~\ref{e0_4.e-10_long}, the solid line plots the results
of the simulation and the dashed line shows the sinusoidal fit used to
derive the turbulent resistivity and magnetic Prandtl number. As expected,
figure~\ref{mean_bz_vs_x} shows that the amplitude of the magnetic field
increases with the forcing amplitude. It also illustrates the quality
of the sinusoidal fit for these models. The turbulent resistivity and
magnetic Prandtl numbers we calculate using the amplitude of the steady-state
field are shown in Table~\ref{table_forcing}. We also report (columns
4 to 7) the time averaged $\alpha$ values and time averaged velocity
fluctuations measured in these runs. As forcing is increased, we find
that $\eta_{t,x}$ increases.  Increasing the forcing by a a factor of
$20$ increases $\eta_{t,x}$ by about $2.5$.  At the same time, $\alpha$
roughly triples. This leads to a mild but systematic increase in $Pm_t$
with the amplitude of the forced vertical magnetic field. To further test
our results, we also computed a model with $|\bb{E}|=4 \times 10^{-9}$
with Athena (model {\it Ey-4E-9-A} in Table~\ref{table_forcing}). The
results are seen to be consistent with the results obtained with Ramses.

Since Eq.~(\ref{soca_eq}) predicts the scaling of $\eta_{t,x}$ with the
velocity fluctuations, in figure~\ref{eta_x_plot} we show our
results in the $(\delta v_x,\eta_{t,x})$ plane. The data 
obtained with Ramses are plotted using diamonds while the points
computed using Athena appear as a star symbol. The dotted line
is a naive estimate that uses the
prediction of the SOCA given by Eq.~(\ref{soca_eq}), assuming that the
correlation timescale is independent of the velocity fluctuations:
$\tau_{corr}=0.15 T_{orb}$. Clearly, the dotted line does not fit the
data. This is most likely because $\tau_{corr}$ varies as the
strength of the turbulence changes. On dimensional grounds, it can be
estimated that
\begin{equation}
\tau_{corr} \sim l_{x,corr}/\delta v_x \, ,
\label{correl_time_eq}
\end{equation}
where $l_{x,corr}$ is a correlation length associated with the
vertical magnetic field fluctuations. Although it is rather
ill--defined, an order of magnitude estimate for $l_{x,corr}$ can be
obtained as \citep{lesur&longaretti07} 
\begin{equation}
l_{x,corr} \sim \left< \frac{\int B_z(x,y,z)B_z(x',y,z)dx dx'}{\int
  B_z^2(x,y,z)dx} \right> \, ,
\label{correl_length_eq}
\end{equation}
where angled brackets denote a spatial average over the $y$
and $z$ directions followed by a time average. Using
Eq.~(\ref{correl_length_eq}), we found that $l_{x,corr}$ is only
weakly varying with the strength of the turbulence. Indeed, we
obtained $0.06 H<l_{x,corr}<0.07 H$ for forcing amplitudes ranging
from $|\bb{E}|=0$ (i.e. no forcing) to $|\bb{E}|=8 \times 10^{-9}$
(i.e. the largest forcing amplitude we considered). It would be
dangerous at this stage to take the estimate given by
Eq.~(\ref{correl_length_eq}) as a good way of calculating
$\tau_{corr}$, as Eq.~(\ref{correl_time_eq}) is only valid on
dimensional grounds. Instead, we used the independence of $l_{x,corr}$
to write the correlation timescale as 
\begin{equation}
\tau_{corr} \sim \tau_0 \frac{\delta v_{x0}}{\delta v_x} \, ,
\end{equation}
where $\tau_0=0.15 T_{orb}$ and $\delta v_{x0}$ denotes the velocity
fluctuations in the absence of forcing (given in
Table~\ref{table_noforcing}). This gives
\begin{equation}
\eta_{t,x}^{SOCA}=\delta v_x \delta v_{x0} \tau_0 \, ,
\end{equation}
which we used to compute the dashed line shown in
figure~\ref{eta_x_plot}. The agreement with the data is much better
and captures the scaling of the turbulent resistivity as the amplitude
of the forcing EMFs is varied by more than one order of magnitude.

\subsection{Radial diffusion of a vertical field in extended boxes}
\label{bigboxes_sec}

Next, we consider the turbulent resistivity in boxes extended in the
radial direction. These models are based on the results of model {\it
A--LB}, which we restarted at time $t=30$ with forcing of
different amplitudes. We only consider forcing in the
azimuthal direction in this case. Again, we report the properties of
the turbulence and the values of $\eta_{t,x}$ and $Pm_t$ we obtained in
Table~\ref{table_forcing}.

To check the results obtained in smaller boxes, we first consider
model {\it Ey-4E-9-A-L4-lx1}. In this case,  we used
$\lambda_{E,x}=H$ and $E_{0y}=4 \times 10^{-10}$. Thus, it is the
big box equivalent of models {\it Ey-4E-9-R} and {\it
Ey-4E-9-A}.  We therefore expect it to yield the same
results. Table~\ref{table_forcing} shows that this is indeed the case,
with the values of $\alpha$, $\eta_{t,x}$ and $Pm_t$ being nearly
the same in all three models.

Next, we turn to models {\it Ey-4E-10-A-L4-lx4} and {\it
Ey-4E-9-A-L4-lx4} for which $\lambda_{E,x}=4H$. The first result to note
is that the amplitude of the vertical field we obtained is larger in
these two cases than in the small box models with equal forcing
EMFs. This was somewhat expected. Since $\lambda_{E,x}$ is now larger,
the local gradient of the vertical field are smaller in the big boxes
and the diffusion of magnetic field is less efficient. Nevertheless, the
measured turbulent resistivity are significantly larger than their
small boxes counterpart. At the same time, the
velocity fluctuations in the radial direction are comparable to the
small boxes models having the same forced vertical field amplitude. 
It is likely that all other statistical properties of the
turbulence share the same insensitivity to box size, which means
that a straightforward application of the SOCA breaks down in this
case. The increased resistivity is therefore due to the increased
wavelength $\lambda_{E,x}$ of the forcing EMF used in that case. One
possible explanation, to be taken with care, for such large $\eta_t$
can then be described as follows: the resulting forced magnetic field,
having a larger scale, will most likely be diffused
by larger eddies. Because the kinetic energy power
spectrum is decreasing with wavenumber, such eddies have a larger
kinetic energy that makes them more efficient at diffusing the
magnetic field.

The consequence of the larger turbulent resistivities obtained in our
large box simulations is lower values of $Pm_t$. In the case of model {\it
Ey-4E-10-A-L4-lx4}, $Pm_t \sim 0.5$, i.e. a factor of about three times
smaller compared to the small boxes.  We recover $Pm_t$ of order unity for
model {\it Ey-4E-9-A-L4-lx4} owing to a larger value of $\alpha$. This is
presumably due to the fact that the flow starts to behave as if threaded
by a net vertical flux for this long wavelength, large amplitude forcing.

\subsection{Vertical diffusion of an azimuthal magnetic field}
\label{vert_diffusion}

In addition to studying radial diffusion of vertical magnetic fields
as described in the previous section, we also performed simulations
in which we study the vertical diffusion of an azimuthal magnetic
field. In this case, 
we restricted our analysis to small boxes, and we chose
$E_{0y}=0$ and therefore considered forcing EMFs purely in the radial
direction. We allowed for variations of the amplitude of the
forcing in the range $4 \times 10^{-9}<E_{0x}< 8 \times 10^{-9}$. We
also tried $E_{0x}=4 \times 10^{-10}$ as in the case of radial
diffusion. However, because the azimuthal magnetic field fluctuations
are of larger amplitudes than the vertical field fluctuations, it has
proven difficult to extract a meaningful mean field from that
model. We therefore decided to consider only large amplitudes for the
forcing EMFs.

The results we obtained are summarized once more in
Table~\ref{table_forcing}. We obtain turbulent resisitivities that are
generally lower than for the case of radial field diffusion. This is
not inconsistent with the prediction of SOCA, as the velocity
fluctuations in the vertical direction are of lower amplitudes than
the velocity fluctuations in the radial direction. Using 
\begin{equation}
\eta_{t,z}^{SOCA}=\delta v_z^2 \tau_{corr} \, ,
\label{soca_vert_eq}
\end{equation}
we obtained $\eta_{t,x}^{SOCA}=6.8 \times 10^{-6}$ for model {\it
Ex-4E-9-R}, while Table~\ref{table_forcing} gives
$\eta_{t,z}=1.05 \times 10^{-5}$ (as compared to $\eta_{t,x}=1.81
\times 10^{-5}$ for model {\it Ey-4E-9-R}). Although it is less accurate than
the estimates we obtained in the case of radial field diffusion, the
prediction of the SOCA still gives a reasonable value for the
turbulent resistivity.

Finally, because of the lower values of the turbulent resistivity,
the final column of Table~\ref{table_forcing} reports systematically
larger values of the turbulent magnetic Prandtl number, 
ranging from $1.57$ for model {\it Ex-4E-9-R} to $3.00$ for model
{\it Ex-4E-9-R}. As in the case of a vertical field diffusing
radially, $Pm_t$ is a slowly increasing function of the strength of
the turbulence.

\section{Discussion and conclusion}
\label{conclusion_sec}

In this paper, we have presented a set of local numerical simulations
aimed at measuring the properties of turbulent diffusion of magnetic
field, quantified using an anomalous turbulent resistivity $\eta_t$,
resulting from MHD turbulence induced by the MRI.  We considered only
the case in which turbulence develops in the absence of a mean magnetic
field. We used two different codes, Athena and Ramses, and vary both the
box size and the magnetic field geometry.  The main result that emerges
from our simulations is that $\eta_t$ tends to be slightly smaller,
but similar in magnitude, to the anomalous viscosity $\nu_t$ that can
be associated with the outward transport of angular momentum due to the
turbulence. In other words, their ratio, the turbulent magnetic number
$Pm_t$, is of order unity (actually in most of our runs it was slightly
larger than one). We also found that our results are roughly consistent
with the predictions of the mean field theory SOCA.  Perhaps more
relevant is the fact that our results are consistent with those recently
published by \citet{guan&gammie09} and \citet{lesur&longaretti09},
who used different field topologies (toroidal and vertical mean field),
different numerical methods (similar to those in the ZEUS code, and an
incompressible pseudo--spectral code respectively) and, as described
in the introduction, different approaches to measure the turbulent
resistivity. This broad agreement demonstrates that all these results,
obtained and published independently, are largely insensitive to the
numerical method and the topogoly of the magnetic field. As suggested
already in the literature \citep{parker71}, a safe first order guess
of the magnitude of the turbulent resistivity is therefore $\eta_t
\sim \nu_t$. 

A number of other results emerge from the comparison of the suite of
simulations we present here with the work of \citet{guan&gammie09}
and \citet{lesur&longaretti09}. First, it is worth noting that 
our results and those of \citet{lesur&longaretti09} were obtained
including explicit dissipation, while those of \citet{guan&gammie09}
relied only on numerical dissipation. Since all the results are roughly
consistent with one another, this is apparently not an issue as far as
determining $Pm_t$ is concerned. Moreover,
  \citet{lesur&longaretti09} consider the case $Pm=1$ while we have used
  $Pm=4$. The broad agreement between both studies therefore suggest
  that $Pm_t$ does not strongly depend on microscopic dissipation
  coefficients. However, it would be premature to draw definite
  statements here since other differences between both works might mask
a possible $Pm$ dependence in the results. Although very
computationally expensive, a systematic study of the
sensitivity of our results to the value of $Pm$ is needed in the
future. Second, $Pm_t$ appears to be smaller in large boxes. Indeed,
we found $Pm_t=0.57$  for model {\it Ey-4E-10-A-L4-lx4}, while $Pm_t$
is found to be larger than 
one in small boxes having identical parameters. This rather low value,
obtained in a large box and for a small forcing EMF, is very close to the
values quoted by \citet{guan&gammie09} for models having the same radial
box size and a weak imposed field. In addition, \citet{guan&gammie09}
considered boxes larger than $4H$ in the radial direction and
found this value of $Pm_t$ to be independent of box size. By contrast,
both our results and those of \citet{lesur&longaretti09} suggest a value
larger than one (or equivalently fairly low values for the turbulent
resisitivity) in small boxes, a regime \citet{guan&gammie09} did not
investigate. Taken together, all these results point 
toward a decrease of $Pm_t$ when box size is increased from values of
a few to values of roughly one half, with a plateau obtained for boxes
larger than $4H$.

Despite their broad agreement, it is instructive to consider the
differences between our results and the aformentionned papers
\citep{guan&gammie09,lesur&longaretti09}. One such difference is that,
regardless of the model or the box size, we always find that $Pm_t$
increases with the amplitude of the forcing
EMF. \citet{lesur&longaretti09} also report such a trend for the
models they label XXZn, in which they consider the vertical diffusion
of a toroidal field, as we do in section~\ref{vert_diffusion}. These
trend is not so clear for their other cases. However, most of their
runs were obtained in the presence of a net vertical flux. The nature
of the turbulence is strongly modified in that case and this is
probably not appropriate to draw the comparison with those models any
further. By contrast, \citet{guan&gammie09} report no dependence of
$Pm_t$ on 
the amplitude of the magnetic field they impose for models that are
similar to ours (i.e. the diffusion of a radially varying vertical
field). Even if the different field topology they consider (namely a
net toroidal flux) plays a role, the difference with our results
can most likely be attributed to the different methods that are being
used. While the present paper aims to measure a steady state
response of the flow to a perturbation that is imposed at all times,
\citet{guan&gammie09} Superpose 
an additional field to an already turbulent flow at $t=0$ and let it
decay. Thus it is possible that transients associated with
this additional field may complicate the estimate of time
averaged values for the transport coefficients\footnote{Of course,
such transients are most likely of physical origin, in the sense that
the decay timescale measured by \citet{guan&gammie09} is related to
the relaxation timescale of the turbulence. Nevertheless, 
their presence complicate the estimate of $Pm_t$}. In particular, for
large strength of the additional field, the flow may not have enough
time to reach fully saturated values of the turbulence before the
strength of the imposed field decays significantly. This might
lead to an underestimate of the value of $\alpha$ and consequently of
the value of $Pm_t$, masking the increase of $Pm_t$ with forcing that
we found.  

It is also important to stress that all three approaches share
common limitations to their analyses. First and foremost, the analysis is
by definition local, whereas the diffusion of magnetic field accross the
disk is an intrisically global problem that depends on the large scale
properties of the disk such as the radial profiles of the surface density
and magnetic flux \citep{spruit&uzdensky05}. Another related limitation
is the neglect of vertical density stratification in the disk. Indeed, all
three studies presents numerical simulations performed in ``unstratified''
shearing boxes, neglecting the vertical component of gravity. In
simulations including gas density stratification, it was found that MHD
turbulence is suppressed in the low density corona above and below the
disk midplane \citep{stoneetal96}. It is likely that turbulent diffusion
will be greatly reduced at those locations, possibly causing the magnetic
flux to remain anchored to the disk corona. This could prevent diffusion
that otherwise would have been driven by turbulence in the midplane,
or by channel modes that couple the upper layers of the disk with the
midplane \citep{bisno&lovelace07,rothstein&lovelace08,lovelaceetal09}. A
numerical test of this scenario is beyong the scope of the present paper
but should be considered in the future.  Finally, let us
  mention the very recent work of \citet{beckwithetal09}. Using 
  global simulations that combine the large scale nature of magnetic
  field diffusion and the vertical density stratication (but, of
  course, with a dramatic reduction in the spatial resolution), these
  authors challenge the concept of turbulent resistivity itself and
  advocate a  
completely different scenario to describe magnetic flux evolution in
accretion disks. Clearly, despite the agreement with
already published calculations, the results presented here should be
taken with care when used in the context of astrophysical applications,
such as jet launching or magnetic flux distribution in accretion disks.

\section*{ACKNOWLEDGMENTS}
We thank J. Goodman for suggesting the method used in this paper
to measure the turbulent resistivity,
and G. Lesur and P.-Y. Longaretti for stimulating discussions. SF
acknowledges the Astronomy Department of Princeton University and the
Institute of Advanced Studies for their hospitality during visits that
enabled this work to be initiated.
This work used the computational facilities supported by the Princeton
Institute for Computational Science and Engineering.

\bibliographystyle{aa}
\bibliography{author}

\newcommand{\noopsort}[1]{}
\begin{thebibliography}{48}
\expandafter\ifx\csname natexlab\endcsname\relax\def\natexlab#1{#1}\fi

\bibitem[{Balbus \& Hawley(1991)}]{balbus&hawley91}
Balbus, S. \& Hawley, J. 1991, ApJ, 376, 214

\bibitem[{Balbus \& Hawley(1998)}]{balbus&hawley98}
Balbus, S. \& Hawley, J. 1998, Rev.Mod.Phys., 70, 1

\bibitem[{{Beckwith} {et~al.}(2009){Beckwith}, {Hawley}, \&
  {Krolik}}]{beckwithetal09}
{Beckwith}, K., {Hawley}, J.~F., \& {Krolik}, J.~H. 2009, ArXiv e-prints

\bibitem[{{Bisnovatyi-Kogan} \& {Lovelace}(2007)}]{bisno&lovelace07}
{Bisnovatyi-Kogan}, G.~S. \& {Lovelace}, R.~V.~E. 2007, \apjl, 667, L167

\bibitem[{{Blandford} \& {Payne}(1982)}]{blandford&payne82}
{Blandford}, R.~D. \& {Payne}, D.~G. 1982, \mnras, 199, 883

\bibitem[{{Brandenburg} \& {Subramanian}(2005)}]{brandenburg&subramanian05}
{Brandenburg}, A. \& {Subramanian}, K. 2005, \physrep, 417, 1

\bibitem[{{Casse} \& {Ferreira}(2000)}]{casse&ferreira00}
{Casse}, F. \& {Ferreira}, J. 2000, \aap, 353, 1115

\bibitem[{{Fromang} {et~al.}(2006){Fromang}, {Hennebelle}, \&
  {Teyssier}}]{Fromangetal06}
{Fromang}, S., {Hennebelle}, P., \& {Teyssier}, R. 2006, \aap, 457, 371

\bibitem[{{Fromang} \& {Nelson}(2009)}]{fromang&nelson09}
{Fromang}, S. \& {Nelson}, R.~P. 2009, \aap, 496, 597

\bibitem[{{Fromang} \& {Papaloizou}(2006)}]{fromang&pap06}
{Fromang}, S. \& {Papaloizou}, J. 2006, A\&A, 452, 751

\bibitem[{{Fromang} {et~al.}(2007){Fromang}, {Papaloizou}, {Lesur}, \&
  {Heinemann}}]{fromangetal07}
{Fromang}, S., {Papaloizou}, J., {Lesur}, G., \& {Heinemann}, T. 2007, \aap,
  476, 1123

\bibitem[{{Gammie}(2001)}]{gammie01}
{Gammie}, C.~F. 2001, ApJ, 553, 174

\bibitem[{{Gardiner} \& {Stone}(2005{\natexlab{a}})}]{gardiner&stone05a}
{Gardiner}, T.~A. \& {Stone}, J.~M. 2005{\natexlab{a}}, Journal of
  Computational Physics, 205, 509

\bibitem[{{Gardiner} \& {Stone}(2005{\natexlab{b}})}]{gardiner&stone05b}
{Gardiner}, T.~A. \& {Stone}, J.~M. 2005{\natexlab{b}}, in AIP Conf. Proc. 784:
  Magnetic Fields in the Universe: From Laboratory and Stars to Primordial
  Structures., ed. E.~M. {de Gouveia dal Pino}, G.~{Lugones}, \& A.~{Lazarian},
  475--488

\bibitem[{{Gardiner} \& {Stone}(2008)}]{gardiner&stone2008}
{Gardiner}, T.~A. \& {Stone}, J.~M. 2008, Journal of Computational Physics,
  227, 4123

\bibitem[{{Goldreich} \& {Lynden-Bell}(1965)}]{goldreich&lyndenbell65}
{Goldreich}, P. \& {Lynden-Bell}, D. 1965, MNRAS, 130, 125

\bibitem[{{Guan} \& {Gammie}(2009)}]{guan&gammie09}
{Guan}, X. \& {Gammie}, C.~F. 2009, \apj, 697, 1901

\bibitem[{{Guan} {et~al.}(2009){Guan}, {Gammie}, {Simon}, \&
  {Johnson}}]{guanetal09}
{Guan}, X., {Gammie}, C.~F., {Simon}, J.~B., \& {Johnson}, B.~M. 2009, \apj,
  694, 1010

\bibitem[{Hawley \& Stone(1995)}]{hawley&stone95}
Hawley, J. \& Stone, J. 1995, Comput. Phys. Commun., 89, 127

\bibitem[{{Hawley} {et~al.}(1995){Hawley}, {Gammie}, \&
  {Balbus}}]{hawleyetal95}
{Hawley}, J.~F., {Gammie}, C.~F., \& {Balbus}, S.~A. 1995, ApJ, 440, 742

\bibitem[{{Hawley} {et~al.}(1996){Hawley}, {Gammie}, \&
  {Balbus}}]{hawleyetal96}
{Hawley}, J.~F., {Gammie}, C.~F., \& {Balbus}, S.~A. 1996, ApJ, 464, 690

\bibitem[{{Heinemann} \&
  {Papaloizou}(2008{\natexlab{a}})}]{heinemann&papaloizou09a}
{Heinemann}, T. \& {Papaloizou}, J.~C.~B. 2008{\natexlab{a}}, ArXiv e-prints

\bibitem[{{Heinemann} \&
  {Papaloizou}(2008{\natexlab{b}})}]{heinemann&papaloizou09b}
{Heinemann}, T. \& {Papaloizou}, J.~C.~B. 2008{\natexlab{b}}, ArXiv e-prints

\bibitem[{{Johansen} {et~al.}(2008){Johansen}, {Youdin}, \&
  {Klahr}}]{johansenetal09}
{Johansen}, A., {Youdin}, A., \& {Klahr}, H. 2008, ArXiv e-prints

\bibitem[{{Johnson} \& {Gammie}(2005)}]{johnson&gammie05}
{Johnson}, B.~M. \& {Gammie}, C.~F. 2005, \apj, 635, 149

\bibitem[{{Johnson} {et~al.}(2008){Johnson}, {Guan}, \&
  {Gammie}}]{johnsonetal08}
{Johnson}, B.~M., {Guan}, X., \& {Gammie}, C.~F. 2008, \apjs, 177, 373

\bibitem[{{Landau} \& {Lifshitz}(1959)}]{landau&lifchitz59}
{Landau}, L.~D. \& {Lifshitz}, E.~M. 1959, {Fluid mechanics} (Course of
  theoretical physics, Oxford: Pergamon Press, 1959)

\bibitem[{{Latter} {et~al.}(2009){Latter}, {Lesaffre}, \&
  {Balbus}}]{latteretal09}
{Latter}, H.~N., {Lesaffre}, P., \& {Balbus}, S.~A. 2009, \mnras, 394, 715

\bibitem[{{Lesur} \& {Longaretti}(2007)}]{lesur&longaretti07}
{Lesur}, G. \& {Longaretti}, P.-Y. 2007, \mnras, 378, 1471

\bibitem[{{Lesur} \& {Longaretti}(2009)}]{lesur&longaretti09}
{Lesur}, G. \& {Longaretti}, P.-Y. 2009, \mnras, submitted

\bibitem[{{Lovelace} {et~al.}(2009){Lovelace}, {Bisnovatyi-Kogan}, \&
  {Rothstein}}]{lovelaceetal09}
{Lovelace}, R.~V.~E., {Bisnovatyi-Kogan}, G.~S., \& {Rothstein}, D.~M. 2009,
  Nonlinear Processes in Geophysics, 16, 77

\bibitem[{{Lubow} {et~al.}(1994){Lubow}, {Papaloizou}, \&
  {Pringle}}]{lubowetal94}
{Lubow}, S.~H., {Papaloizou}, J.~C.~B., \& {Pringle}, J.~E. 1994, \mnras, 267,
  235

\bibitem[{{Masset}(2000)}]{masset00}
{Masset}, F. 2000, \aaps, 141, 165

\bibitem[{{Parker}(1971)}]{parker71}
{Parker}, E.~N. 1971, \apj, 163, 279

\bibitem[{{Pessah} \& {Goodman}(2009)}]{pessah&goodman09}
{Pessah}, M.~E. \& {Goodman}, J. 2009, ArXiv e-prints

\bibitem[{{Pudritz} {et~al.}(2007){Pudritz}, {Ouyed}, {Fendt}, \&
  {Brandenburg}}]{pudritzetal07}
{Pudritz}, R.~E., {Ouyed}, R., {Fendt}, C., \& {Brandenburg}, A. 2007, in
  Protostars and Planets V, ed. B.~{Reipurth}, D.~{Jewitt}, \& K.~{Keil},
  277--294

\bibitem[{{R{\"a}dler} \& {Rheinhardt}(2007)}]{radler&rheinhardt07}
{R{\"a}dler}, K.-H. \& {Rheinhardt}, M. 2007, Geophysical and Astrophysical
  Fluid Dynamics, 101, 117

\bibitem[{{Rothstein} \& {Lovelace}(2008)}]{rothstein&lovelace08}
{Rothstein}, D.~M. \& {Lovelace}, R.~V.~E. 2008, \apj, 677, 1221

\bibitem[{{Schekochihin} {et~al.}(2007){Schekochihin}, {Iskakov}, {Cowley},
  {McWilliams}, {Proctor}, \& {Yousef}}]{Schekochihin07b}
{Schekochihin}, A.~A., {Iskakov}, A.~B., {Cowley}, S.~C., {et~al.} 2007, New
  Journal of Physics, 9, 300

\bibitem[{{Shakura} \& {Sunyaev}(1973)}]{shakura&sunyaev73}
{Shakura}, N.~I. \& {Sunyaev}, R.~A. 1973, A\&A, 24, 337

\bibitem[{{Simon} {et~al.}(2009){Simon}, {Hawley}, \& {Beckwith}}]{simonetal09}
{Simon}, J.~B., {Hawley}, J.~F., \& {Beckwith}, K. 2009, \apj, 690, 974

\bibitem[{{Spruit} \& {Uzdensky}(2005)}]{spruit&uzdensky05}
{Spruit}, H.~C. \& {Uzdensky}, D.~A. 2005, \apj, 629, 960

\bibitem[{{Stone} \& {Gardiner}(2009)}]{stone&gardiner09}
{Stone}, J.~M. \& {Gardiner}, T.~A. 2009, \apj, in preparation

\bibitem[{{Stone} {et~al.}(2008){Stone}, {Gardiner}, {Teuben}, {Hawley}, \&
  {Simon}}]{stoneetal08}
{Stone}, J.~M., {Gardiner}, T.~A., {Teuben}, P., {Hawley}, J.~F., \& {Simon},
  J.~B. 2008, \apjs, 178, 137

\bibitem[{{Stone} {et~al.}(1996){Stone}, {Hawley}, {Gammie}, \&
  {Balbus}}]{stoneetal96}
{Stone}, J.~M., {Hawley}, J.~F., {Gammie}, C.~F., \& {Balbus}, S.~A. 1996, ApJ,
  463, 656

\bibitem[{{Stone} \& {Norman}(1992)}]{stone&norman92b}
{Stone}, J.~M. \& {Norman}, M.~L. 1992, ApJS, 80, 791

\bibitem[{{Teyssier}(2002)}]{Teyssier02}
{Teyssier}, R. 2002, A\&A, 385, 337

\bibitem[{{van Ballegooijen}(1989)}]{vanballegooijen89}
{van Ballegooijen}, A.~A. 1989, in Astrophysics and Space Science Library, Vol.
  156, Accretion Disks and Magnetic Fields in Astrophysics, ed. G.~{Belvedere},
  99--106

\end{thebibliography}

\end{document}